# The problem of sharp notch in microstructured solids governed by dipolar gradient elasticity

P.A. Gourgiotis ,  M.D. Sifnaiou ,  H.G. Georgiadis *

*Mechanics Division, National Technical University of Athens,*
*Zographou Campus, Zographou, GR-15773, Greece*

**Abstract.**  In this paper, we deal with the asymptotic problem of a body of infinite extent with a notch (re-entrant corner) under remotely applied plane-strain or anti-plane shear loadings. The problem is formulated within the framework of the Toupin-Mindlin theory of dipolar gradient elasticity. This generalized continuum theory is appropriate to model the response of materials with microstructure. A linear version of the theory results by considering a linear isotropic expression for the strain-energy density that depends on strain-gradient terms, in addition to the standard strain terms appearing in classical elasticity. Through this formulation, a microstructural material constant $c$ is introduced, in addition to the standard Lamé constants $(\lambda, \mu)$. The faces of the notch are considered to be traction-free and a boundary-layer approach is followed. The boundary value problem is attacked with the asymptotic Knein-Williams technique. Our analysis leads to an eigenvalue problem, which, along with the restriction of a bounded strain energy, provides the asymptotic fields. The cases of a crack and a half-space are analyzed in detail as limit cases of the general notch (infinite wedge) problem. The results show significant departure from the predictions of the standard fracture mechanics.

---

* Corresponding author. Tel.: +30 210 7721365; fax: +30 210 7721302.
 E-mail address: *georgiad@central.ntua.gr (H.G. Georgiadis)*

# 1. Introduction

The present work is concerned with the determination of the asymptotic displacement, strain and stress fields that develop in the vicinity of the tip of a notch within the framework of the dipolar gradient elasticity. The theory of gradient elasticity was introduced by Toupin (1962) and Mindlin (1964) in an effort to model the mechanical behavior of solids with *microstructure* – see the brief literature review on applications and extensions, below. The basic concept of this general theory lies in the consideration of a medium containing elements or particles (called macro-media), which are in themselves deformable media. This behavior can easily be realized if such a macro-particle is viewed as a collection of smaller sub-particles (called micro-media). In this way, each particle of the continuum is endowed with an *internal* displacement field, which can be expanded as a power series in internal coordinate variables. Within the above context, the lowest-order theory (Toupin-Mindlin theory) is the one obtained by retaining only the first (linear) term of the foregoing series. The general framework also appears under the names 'strain-gradient theory' or 'grade-two theory' or 'dipolar gradient theory'.

In the present study, the most common version of the Toupin-Mindlin theory, i.e. the so-called micro-homogeneous case (see Section 10 in Mindlin, 1964), is employed to deal with the problem of sharp notch. According to this, each material particle has three degrees of freedom (the displacement components – just as in the classical theories) and the micro-density does not differ from the macro-density. Also, among the three forms of that version, we choose form II of Mindlin (1964) which assumes a strain-energy density that is a function of the strain tensor and its gradient. In a way, this form is a *first-step* extension of classical elasticity. We notice that the gradient of strain comprises both rotation and stretch gradients. Therefore, this version of the gradient theory is different from the standard couple-stress theory (Cosserat theory with constrained rotations) assuming a strain-energy density that depends upon the strain and the gradient of rotation (torsion-flexure tensor) only. Also, the dipolar gradient theory is different from the Cosserat (or micropolar) theory that takes material particles with six independent degrees of freedom (three displacement components and three rotation components, the latter involving rotation of a micro-medium w.r.t. its surrounding medium).

An interesting feature of the theory stems from the dependence of the strain energy on the gradient of strain – the new material constants imply the presence of characteristic lengths in the material behavior. These lengths can be related with the size of microstructure. In this way, size effects can be incorporated in the stress analysis in a manner that classical theories cannot afford.

Typical cases of continua amenable to such an analysis are periodic material structures like those, e.g., of crystal lattices, crystallites of a polycrystal or grains of a granular material.

Besides the fundamental papers by Toupin (1962) and Mindlin (1964), important contributions in gradient elasticity are also contained in the works by Green and Rivlin (1964), Bleustein (1967), Mindlin and Eshel (1968), Germain (1973) and Maugin (1979). In a brief literature review now, it should be noticed that the Toupin-Mindlin theory had already some successful applications on stress concentration elasticity problems concerning holes and inclusions, during the sixties and the seventies (see e.g. Cook and Weitsman, 1966; Eshel and Rosenfeld, 1970; 1975). More recently, this approach and related extensions have been employed to analyze various problems involving, among other areas, wave propagation (see e.g. Vardoulakis and Georgiadis, 1997; Georgiadis et al., 2000; Georgiadis et al., 2004), fracture (Wei and Hutchinson, 1997; Zhang et al., 1998; Chen et al., 1998; 1999; Shi et al., 2000b; Georgiadis, 2003; Grentzelou and Georgiadis, 2005; 2008; Radi, 2008; Gourgiotis and Georgiadis, 2009), plasticity (see e.g. Fleck et al., 1994; Vardoulakis and Sulem, 1995; Begley and Hutchinson, 1998; Fleck and Hutchinson, 1997; 1998; Gao et al. 1999; Huang et al., 2000; 2004; Hwang et al., 2002; Radi, 2004; 2007; Gurtin, 2004; Bardella and Giacomini, 2008), mechanics of defects (Lazar and Maugin, 2005; Lazar and Kirchner, 2007), and stress concentration due to discrete loadings (Lazar and Maugin, 2006; Georgiadis and Anagnostou, 2008). In addition, efficient numerical techniques (see e.g. Oden et al. 1970; Shu et al., 1999; Amanatidou and Aravas, 2002; Tsepoura et al., 2002; Engel et al., 2002; Tsamasphyros et al., 2007; Giannakopoulos et al., 2006; Markolefas et al., 2008a; 2008b) have been developed to deal with problems analyzed by the Toupin-Mindlin theory.

Regarding now appropriate length scales for strain gradient theories, it is difficult, in general, to link the material length scales involved in the modeling with specific sizes of the underlying microstructure (see e.g. Gurtin, 2004; Bardella and Giacomini, 2008). On the other hand, Zhang et al. (1998) noted that although strain gradient effects are associated with geometrically necessary dislocations in plasticity, they may also be important for the *elastic* range in microstructured materials. For instance, Chen et al. (1998) developed a continuum model for cellular materials and found that the continuum description of these materials obey a gradient elasticity theory. In the latter study, the intrinsic material length was naturally identified with the cell size. Also, in wave propagation dealing with electronic-device applications, surface-wave frequencies on the order of GHz are often used and therefore wavelengths on the micron order appear (see e.g. White, 1970; Farnell, 1978). In such situations, *dispersion phenomena* at high frequencies can only be explained on the basis of a gradient elasticity theory (Georgiadis et al., 2004). In addition, the latter study provides an estimate for a microstructural parameter (i.e. the so-called gradient coefficient $c$) employed in some simple material models. This was effected by considering that the material is composed wholly of unit cells having the form of cubes with edges of size $2h$ and comparing the forms of dispersion curves of Rayleigh waves obtained by the Toupin-Mindlin approach with the ones obtained by the atomic-lattice analysis of Gazis et al. (1960). It was found that $c$ is of the order of $(0.1h)^2$. Another work that relates the material length scales involved in the modeling with the size of the microstructure is due to Chang et al. (2003). The latter associated the microstructural material constants of the Toupin-Mindlin theory with the particle size, the packing density and the inter-particle stiffness in a granular material. Further, Shi et al. (2000a) have linked the internal constitutive length $\ell$ in the mechanism-based strain gradient plasticity with the Burgers vector $b$. Generally, theories with elastic strain gradient effects are intended to model situations where the intrinsic material lengths are of the order of 0.1 – 10 microns (see e.g. Shi et al., 2000b). Since the strengthening effects arising from strain gradients become important when these gradients are large enough, these effects will be significant when the material is deformed in *very small* volumes, such as in the immediate vicinity of crack tips, notches, small holes and inclusions, and micrometer indentations.

We now focus attention on our specific subject, i.e. the problem of a sharp notch (re-entrant corner) in a body of infinite extent under conditions of plane or anti-plane strain, within form II of gradient elasticity. Notch problems have extensively been studied in the context of classical elasticity. Both analytical and asymptotic techniques have been proposed to explore the nature of the displacement and stress fields in such problems. Some of the earlier contributions are those by Knein (1927), Brahtz (1933) and Williams (1952), who treated the problem of the elastic plane notch (infinite wedge) under various combinations of homogeneous boundary conditions. Other related studies, within classical elasticity, are due to Sternberg and Koiter (1958), Karp and Karal (1962), Neuber (1963), Harrington and Ting (1971), Gregory (1979), Leguillon (1988), Dundurs and Markenscoff (1989), Movchan and Nazarov (1990; 1992), and Morozov and Narbut (1995). A thorough overview on the subject and an extensive list of references can be found in the review article by Sinclair (2004). Finally, we should mention that, within the framework of standard couple-stress elasticity, Bogy and Sternberg (1968) treated the problem of the orthogonal wedge subjected to a distribution of shear tractions and resolved a paradox occurred when the problem is treated by classical elasticity. Incidentally, we note that the standard couple-stress elasticity and form II of gradient elasticity give results for plane-strain boundary value problems that do *not* share the same general features of solution behavior, e.g. order of singularities and crack-face displacements in crack problems (Grentzelou and Georgiadis, 2005; Gourgiotis and Georgiadis, 2008; 2009). This can be realized from the fact that not only the number of traction boundary conditions are different in the two cases (four in form II of gradient theory, three in couple-stress theory) but, also, the governing equations are different.

However, as far as the authors are aware, there are no analytical or numerical results in the literature regarding plane-strain or anti-plane strain notch problems in dipolar gradient elasticity. A few results concern the *limit cases* of a crack (Shi et al., 2000b; Georgiadis, 2003; Grentzelou and Georgiadis, 2005; Georgiadis and Grentzelou, 2006; Wei, 2006; Karlis et al., 2007; Grentzelou and Georgiadis, 2008; Markolefas et al. 2008; Gourgiotis and Georgiadis, 2009) and of a half-space (Lazar and Maugin, 2006; Georgiadis and Anagnostou, 2008).

Here, we treat both notch problems asymptotically (for a general angle $90^o \le a \le 180^o$ of the re-entrant corner) and obtain results that are in agreement with the results of the aforementioned limit cases. Our analysis is based on the Knein-Williams technique (Knein, 1927; Williams, 1952; Karp and Karal 1962; Barber, 1992), according to which a set of $(r, \theta)$ polar coordinates is attached to the tip of the notch and the displacement field is expanded as an asymptotic series of separated variable terms, each satisfying the field equations and the traction-free boundary conditions on the faces of the notch. This procedure leads to an eigenvalue problem, which, along with the restriction of a bounded potential energy, provides the asymptotic displacement, strain and stress fields. The asymptotic results show significant departure from those of classical elasticity: the strain field is always *bounded* (finite) at the vicinity of the tip of the notch. This finding is in agreement with the uniqueness theorem for crack problems in gradient elasticity, where the necessary conditions for uniqueness are bounded displacements *and* strains around a crack tip (Grentzelou and Georgiadis, 2005).

## 2. Basic equations of the dipolar gradient elasticity

Here, we briefly present the basic ideas and equations of form II of Mindlin's theory of dipolar gradient theory of small strains and displacements. For more details, we refer the reader to recent papers by the third author (Georgiadis et al., 2004; Georgiadis and Grentzelou, 2006) and to the fundamental papers by Toupin (1962), Mindlin (1964), Bleustein (1967), and Mindlin and Eshel (1968).

The theory can be introduced by the following form of the first law of thermodynamics

$$\rho \dot{\mathrm{E}} = \tau_{pq}\dot{\varepsilon}_{pq} + m_{rpq}\partial_r\dot{\varepsilon}_{pq} \ , \tag{1}$$

where small strains and displacements are assumed, and a Cartesian rectangular coordinate system $Ox_1x_2x_3$ is considered for a 3D continuum (indicial notation and the summation convention will be used throughout). In the above equation, $\partial_p(\ ) \equiv \partial(\ )/\partial x_p$, a superposed dot denotes time derivative, the Latin indices span the range (1,2,3), $\rho$ is the mass density of the continuum, $\mathrm{E}$ is the internal energy per unit mass, $\varepsilon_{pq} = (1/2)(\partial_p u_q + \partial_q u_p) = \varepsilon_{qp}$ is the linear strain tensor, $u_q$ is the displacement vector, $\tau_{pq}$ is the monopolar stress tensor, and $m_{rpq}$ is the dipolar (or double) stress tensor (a third-rank tensor) expressed in dimensions of [force][length]$^{-1}$. For reference, we also write the definitions of the rotation tensor $\omega_{pq} = (1/2)(\partial_p u_q - \partial_q u_p)$ and the rotation vector $\omega_q = (1/2)e_{qkl}\partial_k u_l$, with $e_{qkl}$ being the Levi-Civita permutation symbol.

The dipolar stress tensor follows from the notion of dipolar forces, which are anti-parallel forces acting between the micro-media contained in the continuum with microstructure. As explained by Green and Rivlin (1964), and Jaunzemis (1967), the notion of multipolar forces arises from a series expansion of the mechanical power containing higher-order velocity gradients.

Next, in accord with (1), the following form is taken for the strain-energy density $W$

$$W \equiv W(\varepsilon_{pq}, \partial_r\varepsilon_{pq}) \ . \tag{2}$$

In what follows, we assume the existence of a positive definite function $W(\varepsilon_{pq}, \partial_r\varepsilon_{pq})$. Of course, (2) allows for non-linear constitutive behavior as well, but in the present study we will confine attention to a linear constitutive law. Further, stresses can be defined in the standard variational manner

$$\tau_{pq} \equiv \frac{\partial W}{\partial \varepsilon_{pq}} \ , \qquad m_{rpq} \equiv \frac{\partial W}{\partial(\partial_r\varepsilon_{pq})} \ , \tag{3a,b}$$

where the following symmetries for the monopolar and dipolar stress tensors are noticed: $\tau_{pq} = \tau_{qp}$ and $m_{rpq} = m_{rqp}$.

Then, the equations of equilibrium (global equilibrium) and the traction boundary conditions along a boundary (local equilibrium) can be obtained from variational considerations (Mindlin 1964; Bleustein, 1967). Assuming the absence of body forces, the appropriate expression of the Principle of Virtual Work is written as (Bleustein, 1967)

$$\int_V \left[\tau_{pq}\delta\varepsilon_{pq} + m_{rpq}\delta(\partial_r\varepsilon_{pq})\right]dV = \int_S t_q^{(n)}\delta u_q\, dS + \int_S T_{qr}^{(n)}\partial_q(\delta u_r)dS \ , \tag{4}$$

where the symbol $\delta$ denotes weak variations and it acts on the quantity existing on its right. In the above equation, $t_q^{(n)}$ is the *true* force surface traction, $T_{pq}^{(n)}$ is the *true* double force surface traction, and $n_p$ is the outward unit normal to the boundary along a section inside the body or along the surface of it. In the present study, we generally assume the absence of body forces, in which case the equations of equilibrium and the traction boundary conditions take the following form

$$\partial_p(\tau_{pq} - \partial_r m_{rpq}) = 0 \quad \text{in} \quad V \ , \tag{5}$$

$$P_q^{(n)} = n_p(\tau_{pq} - \partial_r m_{rpq}) - D_p(n_r m_{rpq}) + (D_j n_j)n_r n_p m_{rpq} \quad \text{on} \quad bdy \ , \tag{6}$$

$$R_q^{(n)} = n_r n_p m_{rpq} \quad \text{on} \quad bdy \ , \tag{7}$$

$$E_q = \left[n_r k_p m_{rpq}\right] \quad \text{on} \quad edge\ C \ , \tag{8}$$

where $V$ is the region (open set) occupied by the body, $bdy$ denotes any boundary along a section inside the body or along the surface of it, $D_p(\ ) \equiv \partial_p(\ ) - n_p D(\ )$ is the surface gradient operator, $D(\ ) \equiv n_p \partial_p(\ )$ is the normal gradient operator. Further, $C$ denotes every edge formed by the intersection of two portions, say $S_1$ and $S_2$ of the (closed) bounding surface $S$, and the double brackets $[\ ]$ indicate that the enclosed quantity is the difference between the values on $S_1$ and $S_2$. Also, the vector **k** is defined as $k_q = e_{rpq}s_r n_p$, where $s_q$ is the unit tangent vector to the curve $C$. Furthermore, according to Bleustein (1967), $P_q^{(n)} \equiv t_q^{(n)} + (D_r n_r)n_p T_{pq}^{(n)} - D_p T_{pq}^{(n)}$ is the auxiliary force traction, $R_q^{(n)} \equiv n_p T_{pq}^{(n)}$ is the auxiliary double force traction and $E_q \equiv \left[k_p T_{pq}\right]$ is a line load defined on the edge $C$.

Introducing the constitutive equations of the theory is now in order. The simplest possible linear and isotropic equations result from the following strain-energy density function (Georgiadis et al., 2004; Lazar and Maugin, 2005)

$$W = (1/2)\lambda \varepsilon_{pp}\varepsilon_{qq} + \mu \varepsilon_{pq}\varepsilon_{pq} + (1/2)\lambda c(\partial_r \varepsilon_{pp})(\partial_r \varepsilon_{qq}) + \mu c(\partial_r \varepsilon_{pq})(\partial_r \varepsilon_{pq}),  \qquad (9)$$

where $c$ is the gradient coefficient having dimensions of [length]$^2$, and $(\lambda, \mu)$ are the standard Lamé constants with dimensions of [force][length]$^{-2}$. In this way, only one new material constant is introduced with respect to classical linear isotropic elasticity. Combining (3) with (9) provides the following constitutive equations

$$\tau_{pq} = \lambda \delta_{pq}\varepsilon_{jj} + 2\mu \varepsilon_{pq}, \quad m_{rpq} = c\partial_r(\lambda \delta_{pq}\varepsilon_{jj} + 2\mu \varepsilon_{pq}) = c\partial_r \tau_{pq}, \qquad (10a,b)$$

where $\delta_{pq}$ is the Kronecker delta.

As Lazar and Maugin (2005) pointed out, the particular choice of (9) is physically justified and possesses a notable symmetry. To expose this symmetry, we first consider the general expression (definition) of the strain-energy density $W \equiv \int_0^{\varepsilon_{pq}} \tau_{pq} d\varepsilon_{pq} + \int_0^{\partial_r \varepsilon_{pq}} m_{rpq} d(\partial_r \varepsilon_{pq})$, which for a linear constitutive law takes the form $W = (1/2)\tau_{pq}\varepsilon_{pq} + (1/2)m_{rpq}\partial_r \varepsilon_{pq}$. Then, by virtue of (10b) the strain-energy density in (9) takes the form $W = (1/2)\tau_{pq}\varepsilon_{pq} + c(1/2)(\partial_r \tau_{pq})(\partial_r \varepsilon_{pq})$, which exhibits symmetry with respect to both strain and standard stress. This simple form of the Toupin-Mindlin dipolar gradient elasticity is therefore a *strain gradient* theory as well as a *stress gradient* theory.

Further, as shown by Georgiadis et al. (2004), the restriction of positive definiteness of $W$ requires the following inequalities for the material constants appearing in the theory: $(3\lambda + 2\mu) > 0$, $\mu > 0$, $c > 0$. In addition, stability for the field equations in the general inertial case was proved and to accomplish this, the condition $c > 0$ is a necessary one (Georgiadis et al., 2004).

In the present study, we formulate the 2D plane strain and anti-plane shear notch problems by considering the expression (9) for the strain-energy density $W$. Combining (5) with (10) leads to the field equations of the problem.

## 3. The notch problem in plane-strain

The geometry of the two notch problems in plane-strain (under symmetric and anti-symmetric loading) is shown in Fig. 1. For convenience in the analysis, we will use polar coordinates $(r,\theta)$ with orthonormal base vectors $(\mathbf{e}_r, \mathbf{e}_\theta)$. The faces of the notch are taken along the planes $\theta = \pm a$ ($\mathbf{n} = \pm \mathbf{e}_\theta$) and are assumed to be traction-free.

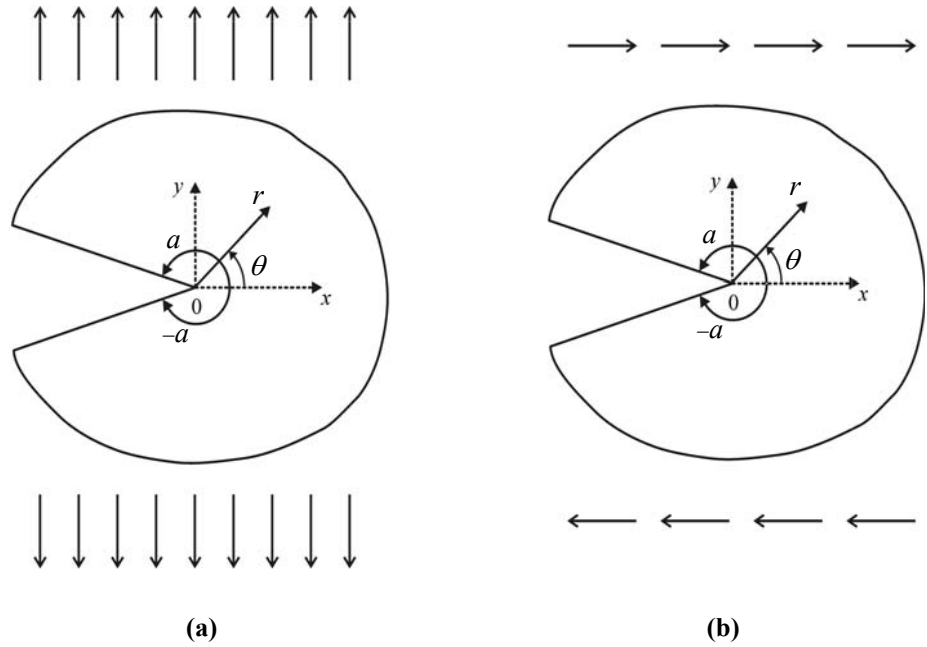

**Fig. 1** The notch problem in plane-strain:
(a) symmetric loading, and (b) anti-symmetric loading.

The displacement field takes the following general form

$$u_r \equiv u_r(r,\theta) \neq 0 , \qquad u_\theta \equiv u_\theta(r,\theta) \neq 0 , \qquad u_z \equiv 0 , \tag{11}$$

whereas the stresses are written as (cf. (10))

$$\tau_{rr} = (\lambda + 2\mu)\partial_r u_r + \lambda r^{-1}(u_r + \partial_\theta u_\theta) , \qquad \tau_{\theta\theta} = (\lambda + 2\mu)r^{-1}(u_r + \partial_\theta u_\theta) + \lambda \partial_r u_r ,$$

$$\tau_{r\theta} = \mu r^{-1}(\partial_\theta u_r - u_\theta) + \mu \partial_r u_\theta , \tag{12a-c}$$

$$m_{rrr} = c\partial_r \tau_{rr}, \quad m_{rr\theta} = c\partial_r \tau_{r\theta}, \quad m_{r\theta\theta} = c\partial_r \tau_{\theta\theta}, \quad m_{\theta rr} = cr^{-1}(\partial_\theta \tau_{rr} - 2\tau_{r\theta}),$$
$$m_{\theta\theta r} = cr^{-1}(\partial_\theta \tau_{\theta r} + \tau_{rr} - \tau_{\theta\theta}), \quad m_{\theta\theta\theta} = cr^{-1}(\partial_\theta \tau_{\theta\theta} + 2\tau_{r\theta}), \tag{13a-f}$$

where $\partial_r(\ ) \equiv \partial(\ )/\partial r$ and $\partial_\theta(\ ) \equiv \partial(\ )/\partial \theta$.

Further, we introduce the so-called total stresses $(t_{\theta r}, t_{\theta\theta})$ stemming from the monopolar traction boundary conditions (see e.g. Georgiadis and Grentzelou, 2006; Gourgiotis and Georgiadis, 2009). To define them, we consider the plane $(r, \theta = \text{const.})$. The normal unit vector to this plane is given as $\mathbf{n} = \mathbf{e}_\theta$. Then, the total stresses *along this plane* are defined, in polar coordinates, as

$$P_r^{(n)} \equiv t_{\theta r} = \tau_{\theta r} - \frac{\partial m_{\theta rr}}{\partial r} - \frac{\partial m_{r\theta r}}{\partial r} - \frac{1}{r}\frac{\partial m_{\theta\theta r}}{\partial \theta} - \frac{1}{r}m_{\theta rr} - \frac{1}{r}m_{r\theta r} + \frac{1}{r}m_{\theta\theta\theta}, \tag{14}$$

$$P_\theta^{(n)} \equiv t_{\theta\theta} = \tau_{\theta\theta} - \frac{\partial m_{\theta\theta r}}{\partial r} - \frac{\partial m_{r\theta\theta}}{\partial r} - \frac{1}{r}\frac{\partial m_{\theta\theta\theta}}{\partial \theta} - \frac{1}{r}m_{r\theta\theta} - \frac{2}{r}m_{\theta\theta r}. \tag{15}$$

The details for the derivation of (14) and (15) is given in Appendix A. These quantities are used to express certain boundary conditions, but they do not possess tensor properties.

Finally, substituting (12) and (13) in (5) leads to the following system of coupled PDEs of the fourth order for the displacement components

$$s_r - c\left(\nabla^2 s_r - r^{-2} s_r - 2r^{-2}\partial_\theta s_\theta\right) = 0, \tag{16a}$$

$$s_\theta - c\left(\nabla^2 s_\theta - r^{-2} s_\theta + 2r^{-2}\partial_\theta s_r\right) = 0. \tag{16b}$$

In the above equations, $\nabla^2(\ ) \equiv \partial_r^2(\ ) + r^{-1}\partial_r(\ ) + r^{-2}\partial_\theta^2(\ )$ is the 2D Laplace operator, and the quantities $(s_r, s_\theta)$ are given by

$$s_r = 2(1-\nu)\partial_r(\partial_r u_r + r^{-1}\partial_\theta u_\theta + r^{-1} u_r) - (1-2\nu)r^{-1}\partial_\theta(\partial_r u_\theta - r^{-1}\partial_\theta u_r + r^{-1} u_\theta), \tag{17a}$$

$$s_\theta = 2(1-\nu)r^{-1}\partial_\theta(\partial_r u_r + r^{-1}\partial_\theta u_\theta + r^{-1} u_r) + (1-2\nu)\partial_r(\partial_r u_\theta - r^{-1}\partial_\theta u_r + r^{-1} u_\theta), \tag{17b}$$

where $\nu = \lambda/2(\lambda + \mu)$ is the Poisson's ratio. The details of the derivation of (16) are also given in Appendix A.

As expected, in the limit $c \to 0$ the Navier-Cauchy equations of classical linear isotropic elasticity are recovered from (16). Indeed, the fact the latter equations have an increased order w.r.t. their limit case (recall that the Navier-Cauchy equations are PDEs of the second order) and the coefficient $c$ multiplies the higher-order term reveal the singular-perturbation character of the gradient theory and the emergence of associated *boundary-layer* effects.

According to the Knein-Williams asymptotic technique in classical elasticity (Knein, 1927; Barber, 1992) and to the HRR field in classical and strain-gradient plasticity (Hutchinson, 1968; Rice and Rosengren, 1968; Chen et al., 1999), an asymptotic expansion of the displacement field is attempted in the form

$$u_r(r,\theta) = r^p U_r^{(1)}(\theta) + r^s U_r^{(2)}(\theta) + ... \ , \quad u_\theta(r,\theta) = r^p U_\theta^{(1)}(\theta) + r^s U_\theta^{(2)}(\theta) + ... \ , \qquad (18a,b)$$

where $(p,s)$ are (in general) complex numbers and $U_r^{(b)}(\theta)$ and $U_\theta^{(b)}(\theta)$ ($b=1,2$) are angular functions.

Now, if the first terms in the above expansions are to be singled out as the dominant ones, $\text{Re}(p) < \text{Re}(s)$, etc. Our search will be restricted only to the dominant terms in such expansions which, accordingly, will give the most singular solution and, thus, the dominant behavior of the stress fields as $r \to 0$.

The boundary conditions for the traction-free notch at $\theta = \pm a$ read

$$t_{\theta\theta}(r,\pm a) = 0 \ , \quad t_{\theta r}(r,\pm a) = 0 \ , \quad m_{\theta\theta r}(r,\pm a) = 0 \ , \quad m_{\theta\theta\theta}(r,\pm a) = 0 \ . \qquad (19a\text{-}d)$$

Further, if only the dominant singular terms are retained in the asymptotic fields, the governing equations in (16) become

$$\nabla^2 s_r - r^{-2} s_r - 2 r^{-2} \partial_\theta s_\theta = 0 \ , \qquad (20a)$$

$$\nabla^2 s_\theta - r^{-2} s_\theta + 2 r^{-2} \partial_\theta s_r = 0 \ . \qquad (20b)$$

The general solution to (20) is obtained as

$$u_r = r^p \left[ A_1 \cos(p-1)\theta + A_2 \cos(p+1)\theta + A_3 \cos(p-3)\theta \right]$$
$$+ r^p \left[ B_1 \sin(p-1)\theta + B_2 \sin(p+1)\theta + B_3 \sin(p-3)\theta \right] \ , \qquad (21)$$

$$u_\theta = r^p \left[ -A_2 \sin(p+1)\theta - A_3 \frac{(p+5-8\nu)}{(p-7+8\nu)} \sin(p-3)\theta + A_4 \sin(p-1)\theta \right]$$

$$+ r^p \left[ B_2 \cos(p+1)\theta + B_3 \frac{(p+5-8\nu)}{(p-7+8\nu)} \cos(p-3)\theta + B_4 \cos(p-1)\theta \right], \qquad (22)$$

where the unknown constants $A_b$ and $B_b$ (with $b = 1,2,3,4$), correspond to symmetric and anti-symmetric loadings, respectively.

Next, we utilize the constitutive equations in (12) and (13), retain only the most singular terms and write the boundary conditions in terms of displacements

$$t_{\theta r}(r, \pm a) = 0 \;\Rightarrow\; -\partial_r m_{\theta rr} - \partial_r m_{r\theta r} - r^{-1}\partial_\theta m_{\theta\theta r} - r^{-1}m_{\theta rr} - r^{-1}m_{r\theta r} + r^{-1}m_{\theta\theta\theta} = 0 \;\Rightarrow$$

$$\Rightarrow -(3-4\nu)r^2 \partial_r^2 \partial_\theta u_r + (3-2\nu)\partial_\theta u_r + (5-6\nu)\partial_\theta^2 u_\theta - r\partial_\theta^2 \partial_r u_\theta +$$

$$+ (1-2\nu)\left[ -r^3 \partial_r^3 u_\theta + 2r^2 \partial_r^2 u_\theta + r\partial_r \partial_\theta u_r - r\partial_r u_\theta - \partial_\theta^3 u_r + u_\theta \right] = 0, \qquad (23)$$

$$t_{\theta\theta}(r, \pm a) = 0 \;\Rightarrow\; -\partial_r m_{\theta\theta r} - \partial_r m_{r\theta\theta} - r^{-1}\partial_\theta m_{\theta\theta\theta} - r^{-1}m_{r\theta\theta} - 2r^{-1}m_{\theta\theta r} = 0 \;\Rightarrow$$

$$\Rightarrow (3-4\nu)r^2 \partial_r^2 \partial_\theta u_\theta + 2(2-3\nu)\partial_\theta^2 u_r + 2\nu r^3 \partial_r^3 u_r + 2\nu \partial_\theta u_\theta + r\partial_\theta^2 \partial_r u_r +$$

$$+ (1-\nu)\left[ 4r^2 \partial_r^2 u_r + 2\partial_\theta^3 u_\theta - 2r\partial_r \partial_\theta u_\theta - 2r\partial_r u_r + 2u_r \right] = 0, \qquad (24)$$

$$m_{\theta\theta r}(r, \pm a) = 0 \;\Rightarrow\; \partial_r \left( r^{-1}(2u_r + \partial_\theta u_\theta) \right) + r^{-2}\partial_\theta (\partial_\theta u_r - 2u_\theta) = 0, \qquad (25)$$

$$m_{\theta\theta\theta}(r, \pm a) = 0 \;\Rightarrow\; (1-\nu)\left( \partial_r (r^{-1}u_\theta) + r^{-2}\partial_\theta (2u_r + \partial_\theta u_\theta) \right) + \nu \partial_r \left( r^{-1}(\partial_\theta u_r - u_\theta) \right) = 0. \quad (26)$$

Now, Eqs. (23)-(26) together with (21) and (22) form an *eigenvalue* problem. For the existence of a non-trivial solution, the determinant of the coefficients of $(A_b, B_b)$ should vanish and this gives the following equations for $p$:

*Symmetric loading*

$$(p-1)^4(p-2)^2\left\{(p-1)\left[2(p-1)(\cos 2(p-1)a+\cos 4a-1)-(p-2)\cos 2(p+1)a\right.\right.$$

$$\left.\left.-p\cos 2(p-3)a\right]+2(5-4v)(1-\cos 4(p-1)a)\right\}=0, \quad (27a)$$

*Anti-symmetric loading*

$$(p-1)^4(p-2)^2\left\{(p-1)\left[2(p-1)(\cos 2(p-1)a-\cos 4a+1)-(p-2)\cos 2(p+1)a\right.\right.$$

$$\left.\left.-p\cos 2(p-3)a\right]-2(5-4v)(1-\cos 4(p-1)a)\right\}=0. \quad (27b)$$

It is worth noting that, in gradient elasticity, the eigenvalues $p$ depend, in general, not only upon the angle of the notch, as in classical elasticity, but also upon the Poisson's ratio $v$. Further, we note that the transcendental equations (27a) and (27b) exhibit an infinite number of solutions. However, not all of them are allowed by the so-called energy criterion (see e.g. Barber, 1992, for the respective problem of classical elasticity). More specifically, in our analysis we consider the body under remotely applied loading, without any concentrated load applied inside the body or on the boundary. Therefore, the total strain-energy in a small region surrounding the notch apex (as $r \to 0$) should vanish. This restricts the number of *acceptable* eigenvalues. It can further be checked that the potential energy $U$ per unit length (along the $z$-axis) in a small circular area around the tip of the notch is given by $U = \int_{-a}^{a}\int_{0}^{r_0} W\, r\, dr\, d\theta$ (Barber, 1992). In our case, it derives from (9) that the strain-energy density $W$ behaves at most as $W \approx \left(\partial \varepsilon_{pq}/\partial r\right)^2 \approx \left(\partial^2 u_p/\partial r^2\right)^2 \approx r^{2p-4}$. Consequently, the potential energy can be written in the form $U = (\text{const.})\int_{0}^{r_0} r^{2p-3}\, dr\, d\theta$, which is *bounded* in the vicinity of the corner point, if and only if $2p-3 > -1 \Rightarrow p > 1$.

Nevertheless, it turns out that the eigenvalue $p = 1$ satisfies, as well, the characteristic equations in (27) for all angles $a$. In this case, the field equations in (20) admit the following special solution

$$u_r = r\left[C_1 + C_2 \sin 2\theta + C_3 \cos 2\theta + C_4 \theta + C_5 \theta \sin 2\theta + C_6 \theta \cos 2\theta\right] , \tag{28a}$$

$$u_\theta = r\left[C_2 \cos 2\theta - C_3 \sin 2\theta + C_4 \frac{(5-6\nu)}{2(3-4\nu)} + C_5\left(\theta \cos 2\theta - \frac{1}{2(3-4\nu)}\sin 2\theta\right)\right.$$
$$\left. - C_6\left(\theta \sin 2\theta + \frac{1}{2(3-4\nu)}\cos 2\theta\right) + C_7 \theta + C_8\right] . \tag{28b}$$

The boundary conditions (23)-(26) necessarily imply that $C_4 = C_5 = C_6 = C_7 = 0$. Also, we set $C_8 = 0$ since this term in $u_\theta$ corresponds to a rigid body motion. Now, it can readily be shown that the above displacement field results in a *constant* strain field, and also, that it does *not* contribute to the dipolar stress field (note that $\partial_r \varepsilon_{pq} = 0$, in this case). Therefore, in this special case, the strain-energy density $W$ behaves as in classical elasticity, i.e. $W \approx \varepsilon_{pq}^2 \approx \left(\partial u_p/\partial r\right)^2 \approx r^{2p-2}$. Consequently, the eigenvalue $p = 1$ is a physically admissible eigenvalue since it leads, for all angles $a$, to a bounded potential energy. The existence of the displacement field associated with the eigenvalue $p = 1$ was first noticed by Radi (2008) for a mode III crack in couple-stress elasticity and by Aravas and Giannakopoulos (2009) for the plane-strain modes of fracture in dipolar gradient elasticity. Finally, we note that the case $p < 1$ is excluded since it *always* leads to an unbounded potential energy.

In light of the above, the displacement field is finally written as

*Symmetric loading*

$$u_r = r\left[C_1 + C_3 \cos 2\theta\right] + r^p\left[A_1 \cos(p-1)\theta + A_2 \cos(p+1)\theta + A_3 \cos(p-3)\theta\right] , \tag{29a}$$

$$u_\theta = -C_3 r \sin 2\theta - r^p\left[A_2 \sin(p+1)\theta + A_3 \frac{(p+5-8\nu)}{(p-7+8\nu)}\sin(p-3)\theta - A_4 \sin(p-1)\theta\right] , \tag{29b}$$

*Anti-symmetric loading*

$$u_r = C_2 r \sin 2\theta + r^p\left[B_1 \sin(p-1)\theta + B_2 \sin(p+1)\theta + B_3 \sin(p-3)\theta\right] , \tag{30a}$$

$$u_\theta = C_2 r \cos\theta + r^p \left[ B_2 \cos(p+1)\theta + B_3 \frac{(p+5-8\nu)}{(p-7+8\nu)} \cos(p-3)\theta + B_4 \cos(p-1)\theta \right], \quad (30b)$$

where $p > 1$. It is noted, that due to the singular perturbation character of the field equations and the boundary conditions, the lower-order terms $(C_1, C_2, C_3)$ are coupled with terms of order $O(r^3)$ to satisfy the conditions of vanishing total stresses at the faces of the notch (Eqs. (19a,b)). Further, it follows from (12)-(15) that the monopolar and dipolar stresses behave as $\sim r^{p-1}$ (bounded variation) and $\sim r^{p-2}$ (singular variation), respectively, whereas the total stresses exhibit the most singular behavior $\sim r^{p-3}$ in the vicinity of the notch apex. The general expressions for the strain and stress fields are given in Appendix B. Next, from the characteristic equations (27a) and (27b), we infer that for angles in the range $90^o \leq a \leq 180^o$ the exponent $p$ is decreasing monotonically.

Now, Fig. 2a depicts the variation of the exponent of the monopolar stress ($p-1$) in both cases of gradient and classical elasticity. We observe, therefore, that the monopolar stress and the strain fields at the apex of a sharp notch are not singular in dipolar gradient elasticity. In addition, Fig. 2b depicts the variation of the exponent of the total stress ($p-3$) in gradient elasticity. It is observed that as the notch angle $a$ is reduced from $180°$ to $90°$, the strength of the singularity falls monotonically from $-1.5$ to $-1$. The most singular eigenvalue occurs in the case of a crack ($a=180°$). In general, the exponent $p$ depends not only upon the notch angle $a$ but also upon the Poisson's ratio $v$. But, the dependence upon $v$ is only marginal.

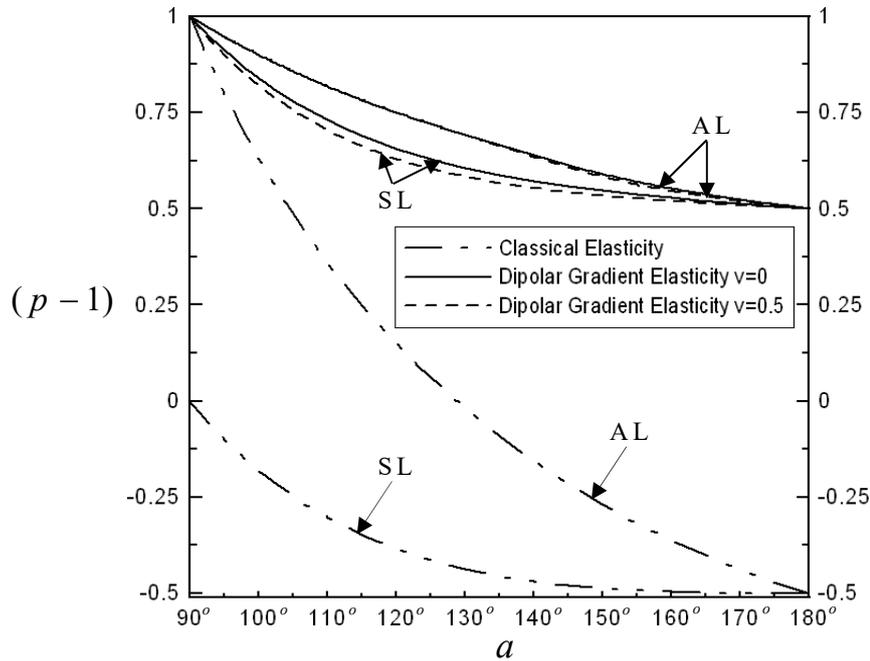

**Fig. 2a** Variation of the exponent of the monopolar stress with respect to the notch angle $a$ for symmetric loading (SL) and anti-symmetric loading (AL).

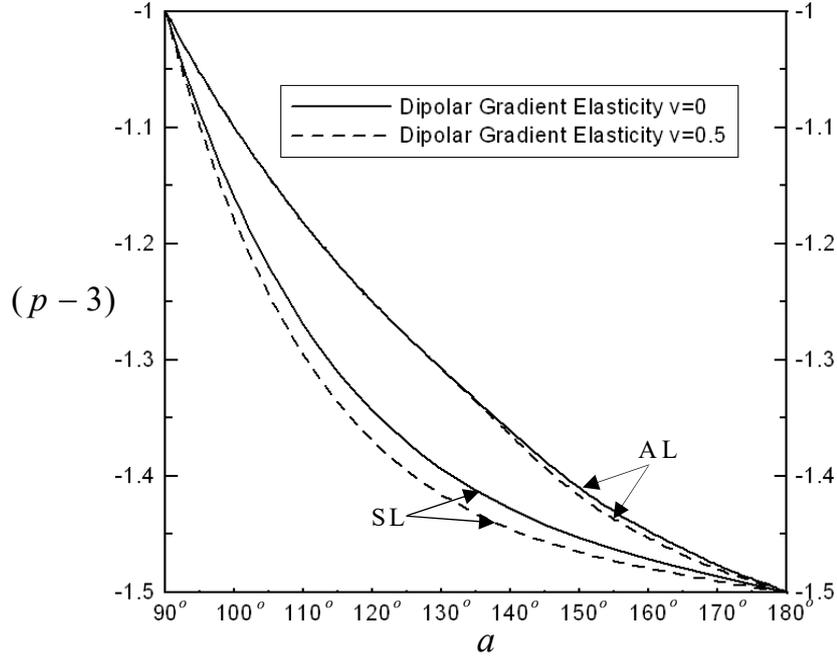

**Fig. 2b** Variation of the exponent of the total stress in gradient elasticity with respect to the notch angle $a$ for symmetric loading (SL) and anti-symmetric loading (AL).

## 4. The limit case of a crack in plane-strain

The limit case $a=180°$, which corresponds to the mode I and mode II crack problems is examined now in detail. In this case, both plane strain modes share the same characteristic equation, viz.

$$(p-1)^4(p-2)^2(1-\cos 4\pi p)=0 \Rightarrow p = n/2 \ , \ n=0,\pm 1,\pm 2,... \ . \tag{31}$$

In this case there is no dependence of the exponent $p$ upon the Poisson's ratio. Then, according to the energy criterion the most singular admissible value of the exponent is $p = 3/2$. Also, from Eqs. (29) and (30), it is clear that the case $p = 1$ should also be taken into account.

Below, we present separately the cases of mode I and mode II crack.

*Mode I crack*

In view of the symmetry of the mode I problem, we obtain the following displacement field

$$u_r = r\left[C_1 + C_3 \cos 2\theta\right] + A_1 r^{3/2}\left[(3-8\nu)\cos\frac{\theta}{2} + 3\frac{(11-16\nu)}{(41-32\nu)}\cos\frac{3\theta}{2}\right]$$
$$- A_2 r^{3/2}\left[3\frac{(11-16\nu)}{(41-32\nu)}\cos\frac{3\theta}{2} - \cos\frac{5\theta}{2}\right], \quad (32a)$$

$$u_\theta = -C_3 r \sin 2\theta + A_1 r^{3/2}\left[(9-8\nu)\sin\frac{\theta}{2} - 3\frac{(13-16\nu)}{(41-32\nu)}\sin\frac{3\theta}{2}\right]$$
$$+ A_2 r^{3/2}\left[3\frac{(13-16\nu)}{(41-32\nu)}\sin\frac{3\theta}{2} - \sin\frac{5\theta}{2}\right], \quad (32b)$$

where $(C_1, C_3)$ are amplitude factors for the lower-order crack-tip fields and $(A_1, A_2)$ are the amplitude factors for the dominant term of order $3/2$. It is noted, that the lower-order terms do not contribute to the energy release rate as well as to the crack-tip opening displacement (Gourgiotis and Georgiadis, 2009). However, these terms produce a constant strain field at the crack tip. A detailed discussion about the significance of these terms and their connection to fracture criteria can be found in Aravas and Giannakopoulos (2009).

Further, by virtue of (32), it is found that the strains and the monopolar stresses are bounded at the crack-tip region

$$\varepsilon_{rr} = C_1 + C_3 \cos 2\theta + \frac{3}{2} A_1 r^{1/2} \left[ \frac{(33-48\nu)}{41-32\nu} \cos \frac{3\theta}{2} + (3-8\nu) \cos \frac{\theta}{2} \right]$$

$$- \frac{3}{2} A_2 r^{1/2} \left[ \frac{(33-48\nu)}{41-32\nu} \cos \frac{3\theta}{2} - \cos \frac{5\theta}{2} \right], \tag{33a}$$

$$\varepsilon_{\theta\theta} = C_1 - C_3 \cos 2\theta - \frac{3}{2} A_1 r^{1/2} \left[ \frac{(17-16\nu)}{41-32\nu} \cos \frac{3\theta}{2} - (5-8\nu) \cos \frac{\theta}{2} \right]$$

$$+ \frac{3}{2} A_2 r^{1/2} \left[ \frac{(17-16\nu)}{41-32\nu} \cos \frac{3\theta}{2} - \cos \frac{5\theta}{2} \right], \tag{33b}$$

$$\varepsilon_{\theta r} = \varepsilon_{r\theta} = -C_3 \sin 2\theta - \frac{3}{2} A_1 r^{1/2} \left[ \frac{(23-32\nu)}{41-32\nu} \sin \frac{3\theta}{2} - \sin \frac{\theta}{2} \right]$$

$$+ \frac{3}{2} A_2 r^{1/2} \left[ \frac{(23-32\nu)}{41-32\nu} \sin \frac{3\theta}{2} - \sin \frac{5\theta}{2} \right], \tag{33c}$$

$$\tau_{rr} = 2\mu C_1/(1-2\nu) + 2\mu C_3 \cos 2\theta + 3\mu A_1 r^{1/2} \left[ 3\cos \frac{\theta}{2} + \frac{(33-32\nu)}{41-32\nu} \cos \frac{3\theta}{2} \right]$$

$$- 3\mu A_2 r^{1/2} \left[ \frac{(33-32\nu)}{41-32\nu} \cos \frac{3\theta}{2} - \cos \frac{5\theta}{2} \right]. \tag{34a}$$

$$\tau_{\theta\theta} = 2\mu C_1/(1-2\nu) - 2\mu C_3 \cos 2\theta + 3\mu A_1 r^{1/2} \left[ 5\cos \frac{\theta}{2} - \frac{(17-32\nu)}{41-32\nu} \cos \frac{3\theta}{2} \right]$$

$$- 3\mu A_2 r^{1/2} \left[ -\frac{(17-32\nu)}{41-32\nu} \cos \frac{3\theta}{2} + \cos \frac{5\theta}{2} \right], \tag{34b}$$

$$\tau_{\theta r} = \tau_{r\theta} = -2\mu C_3 \sin 2\theta + 3\mu A_1 r^{1/2} \left[ \sin \frac{\theta}{2} - \frac{(23-32\nu)}{41-32\nu} \sin \frac{3\theta}{2} \right]$$

$$- 3\mu A_2 r^{1/2} \left[ -\frac{(23-32\nu)}{41-32\nu} \sin \frac{3\theta}{2} + \sin \frac{5\theta}{2} \right]. \tag{34c}$$

Also, the dipolar and the total stresses are written as

$$m_{\theta\theta r} = \frac{3\mu c}{2} A_1 r^{-1/2} \left[ -3\cos\frac{\theta}{2} + \frac{(31-32\nu)}{41-32\nu}\cos\frac{3\theta}{2} \right]$$

$$- \frac{3\mu c}{2} A_2 r^{-1/2} \left[ \frac{(31-32\nu)}{41-32\nu}\cos\frac{3\theta}{2} + \cos\frac{5\theta}{2} \right], \tag{35a}$$

$$m_{\theta\theta\theta} = -\frac{3\mu c}{2} A_1 r^{-1/2} \left[ \sin\frac{\theta}{2} + \sin\frac{3\theta}{2} \right] + \frac{3\mu c}{2} A_2 r^{-1/2} \left[ \sin\frac{3\theta}{2} + \sin\frac{5\theta}{2} \right], \tag{35b}$$

$$m_{rrr} = \frac{3\mu c}{2} A_1 r^{-1/2} \left[ 3\cos\frac{\theta}{2} + \frac{(33-32\nu)}{41-32\nu}\cos\frac{3\theta}{2} \right]$$

$$- \frac{3\mu c}{2} A_2 r^{-1/2} \left[ \frac{(33-32\nu)}{41-32\nu}\cos\frac{3\theta}{2} - \cos\frac{5\theta}{2} \right], \tag{35c}$$

$$m_{r\theta r} = \frac{3\mu c}{2} A_1 r^{-1/2} \left[ \sin\frac{\theta}{2} - \frac{(23-32\nu)}{41-32\nu}\sin\frac{3\theta}{2} \right]$$

$$+ \frac{3\mu c}{2} A_2 r^{-1/2} \left[ \frac{(23-32\nu)}{41-32\nu}\sin\frac{3\theta}{2} - \sin\frac{5\theta}{2} \right], \tag{35d}$$

$$m_{\theta rr} = -\frac{3\mu c}{2} A_1 r^{-1/2} \left[ 7\sin\frac{\theta}{2} + \frac{(7+32\nu)}{41-32\nu}\sin\frac{3\theta}{2} \right]$$

$$+ \frac{3\mu c}{2} A_2 r^{-1/2} \left[ \frac{(7+32\nu)}{41-32\nu}\sin\frac{3\theta}{2} - \sin\frac{5\theta}{2} \right], \tag{35e}$$

$$m_{r\theta\theta} = \frac{3\mu c}{2} A_1 r^{-1/2} \left[ 5\cos\frac{\theta}{2} - \frac{(17-32\nu)}{41-32\nu}\cos\frac{3\theta}{2} \right]$$

$$+ \frac{3\mu c}{2} A_2 r^{-1/2} \left[ \frac{(17-32\nu)}{41-32\nu}\cos\frac{3\theta}{2} - \cos\frac{5\theta}{2} \right], \tag{35f}$$

$$t_{\theta r} = \frac{3\mu c}{4} A_1 r^{-3/2} \left[ \sin\frac{\theta}{2} + \sin\frac{3\theta}{2} \right] - \frac{3\mu c}{4} A_2 r^{-3/2} \left[ \sin\frac{3\theta}{2} + \sin\frac{5\theta}{2} \right], \tag{36a}$$

$$\begin{aligned}t_{\theta\theta} =& \frac{3\mu c}{4} A_1 r^{-3/2} \left[ \frac{(47-32\nu)}{41-32\nu} \cos\frac{3\theta}{2} + 5\cos\frac{\theta}{2} \right] \\ &- \frac{3\mu c}{4} A_2 r^{-3/2} \left[ \frac{(47-32\nu)}{41-32\nu} \cos\frac{3\theta}{2} + \cos\frac{5\theta}{2} \right].\end{aligned} \tag{36b}$$

*Mode II crack*

In view of the anti-symmetry of the mode II problem, we obtain the corresponding displacement field as

$$u_r = C_2 r \sin 2\theta + B_1 r^{3/2} \sin\frac{\theta}{2} + B_2 r^{3/2} \left[ -\frac{3(11-16\nu)}{37-32\nu} \sin\frac{3\theta}{2} + \sin\frac{5\theta}{2} \right], \tag{37a}$$

$$u_\theta = C_2 r \cos 2\theta - B_1 r^{3/2} \cos\frac{\theta}{2} + B_2 r^{3/2} \left[ \cos\frac{5\theta}{2} - \frac{3(13-16\nu)}{37-32\nu} \cos\frac{3\theta}{2} + \frac{12}{37-32\nu} \cos\frac{\theta}{2} \right]. \tag{37b}$$

where the constants $C_2$ and $(B_1, B_2)$ are left unspecified by the asymptotic analysis. Note, that the linear terms in $r$ do not contribute to the crack sliding displacement $u_r$. Indeed, for $\theta = \pm\pi$, the linear term in (37a) vanishes.

Further, the strains and the monopolar, dipolar and total stresses are written as

$$\varepsilon_{rr} = C_2 \sin 2\theta + \frac{3}{2} B_1 r^{1/2} \sin\frac{\theta}{2} - \frac{3}{2} B_2 r^{1/2} \left[ \frac{(33-48\nu)}{37-32\nu} \sin\frac{3\theta}{2} - \sin\frac{5\theta}{2} \right], \tag{38a}$$

$$\begin{aligned}\varepsilon_{\theta\theta} =& -C_2 \sin 2\theta - \frac{3}{2} B_2 r^{1/2} \left[ \frac{4}{37-32\nu} \sin\frac{\theta}{2} - \frac{(17-16\nu)}{37-32\nu} \sin\frac{3\theta}{2} + \sin\frac{5\theta}{2} \right] \\ &+ \frac{3}{2} B_1 r^{1/2} \sin\frac{\theta}{2},\end{aligned} \tag{38b}$$

$$\varepsilon_{r\theta} = \varepsilon_{\theta r} = C_2 \cos 2\theta + \frac{3}{2} B_2 r^{1/2} \left[ \frac{2}{37-32\nu} \cos\frac{\theta}{2} - \frac{(23-32\nu)}{37-32\nu} \cos\frac{3\theta}{2} + \cos\frac{5\theta}{2} \right], \tag{38c}$$

$$\tau_{rr} = 2\mu C_2 \sin 2\theta - 3\mu B_2 r^{1/2} \left[ \frac{4\nu}{(1-2\nu)(37-32\nu)} \sin\frac{\theta}{2} + \frac{(33-32\nu)}{37-32\nu} \sin\frac{3\theta}{2} - \sin\frac{5\theta}{2} \right]$$
$$+ \frac{3\mu}{1-2\nu} B_1 r^{1/2} \sin\frac{\theta}{2}, \tag{39a}$$

$$\tau_{\theta\theta} = -2\mu C_2 \sin 2\theta - 3\mu B_2 r^{1/2} \left[ \frac{4(1-\nu)}{(1-2\nu)(37-32\nu)} \sin\frac{\theta}{2} - \frac{(17-32\nu)}{37-32\nu} \sin\frac{3\theta}{2} + \sin\frac{5\theta}{2} \right]$$
$$+ \frac{3\mu}{1-2\nu} B_1 r^{1/2} \sin\frac{\theta}{2}, \tag{39b}$$

$$\tau_{\theta r} = \tau_{r\theta} = 2\mu C_2 \cos 2\theta + 3\mu B_2 r^{1/2} \left[ \frac{2}{37-32\nu} \cos\frac{\theta}{2} - \frac{(23-32\nu)}{37-32\nu} \cos\frac{3\theta}{2} + \cos\frac{5\theta}{2} \right], \tag{39c}$$

$$m_{\theta\theta\theta} = -\frac{3\mu c}{2} B_2 r^{-1/2} \left[ -\frac{4(1-3\nu)}{(1-2\nu)(37-32\nu)} \cos\frac{\theta}{2} + \frac{(41-32\nu)}{37-32\nu} \cos\frac{3\theta}{2} + \cos\frac{5\theta}{2} \right]$$
$$+ \frac{3\mu c}{2(1-2\nu)} B_1 r^{-1/2} \cos\frac{\theta}{2}, \tag{40a}$$

$$m_{\theta\theta r} = -\frac{3\mu c}{2} B_2 r^{-1/2} \left[ -\frac{6}{37-32\nu} \sin\frac{\theta}{2} + \frac{(31-32\nu)}{37-32\nu} \sin\frac{3\theta}{2} + \sin\frac{5\theta}{2} \right], \tag{40b}$$

$$m_{rrr} = -\frac{3\mu c}{2} B_2 r^{-1/2} \left[ -\sin\frac{5\theta}{2} + \frac{(33-32\nu)}{37-32\nu} \sin\frac{3\theta}{2} + \frac{4\nu}{(37-32\nu)(1-2\nu)} \sin\frac{\theta}{2} \right]$$
$$+ \frac{3\mu c}{2(1-2\nu)} B_1 r^{-1/2} \sin\frac{\theta}{2}, \tag{40c}$$

$$m_{\theta rr} = -\frac{3\mu c}{2} B_2 r^{-1/2} \left[ \frac{4(2-3\nu)}{(1-2\nu)(37-32\nu)} \cos\frac{\theta}{2} + \frac{(7+32\nu)}{37-32\nu} \cos\frac{3\theta}{2} - \cos\frac{5\theta}{2} \right]$$
$$+ \frac{3\mu c}{2(1-2\nu)} B_1 r^{-1/2} \cos\frac{\theta}{2}, \tag{40d}$$

$$m_{r\theta r} = -\frac{3\mu c}{2} B_2 r^{-1/2} \left[ -\frac{2}{37-32\nu} \cos\frac{\theta}{2} + \frac{(23-32\nu)}{37-32\nu} \cos\frac{3\theta}{2} - \cos\frac{5\theta}{2} \right], \tag{40e}$$

$$m_{r\theta\theta} = -\frac{3\mu c}{2} B_2 r^{-1/2} \left[ \frac{4(1-\nu)}{(37-32\nu)(1-2\nu)} \sin\frac{\theta}{2} - \frac{(17-32\nu)}{37-32\nu} \sin\frac{3\theta}{2} + \sin\frac{5\theta}{2} \right]$$
$$+ \frac{3\mu c}{2(1-2\nu)} B_1 r^{-1/2} \sin\frac{\theta}{2}. \tag{40f}$$

$$t_{\theta r} = \frac{3\mu c}{4} B_2 r^{-3/2} \left[ \frac{4(2-5\nu)}{(1-2\nu)(37-32\nu)} \cos\frac{\theta}{2} + \frac{(41-32\nu)}{37-32\nu} \cos\frac{3\theta}{2} + \cos\frac{5\theta}{2} \right]$$
$$+ \frac{3\mu c}{4(1-2\nu)} B_1 r^{-3/2} \cos\frac{\theta}{2}, \tag{41a}$$

$$t_{\theta\theta} = -\frac{3\mu c}{4} B_2 r^{-3/2} \left[ \frac{10}{37-32\nu} \sin\frac{\theta}{2} + \frac{(47-32\nu)}{37-32\nu} \sin\frac{3\theta}{2} + \sin\frac{5\theta}{2} \right]. \tag{41b}$$

Regarding now the previous asymptotic results, we notice the following points:

(i) The total stresses (Eqs. (36), (41)) exhibit a stronger singularity ($\sim r^{-3/2}$) than the one predicted by standard linear fracture mechanics. This type of singularity was also observed in previous studies of crack problems in gradient elasticity (see e.g. Shi et al., 2000b; Georgiadis, 2003; Karlis et al., 2007; Markolefas et al. 2008; Gourgiotis and Georgiadis, 2009). In addition, such a strong singularity was suggested by the experimental evidence of Prakash et al. (1992) in extremely brittle fracture. A previous analytical study (Gourgiotis and Georgiadis, 2009), employing the method of singular integral equations, has shown that these stresses are compressive ahead of the crack-tip exhibiting a cohesive character. The length of this cohesive zone is on the order of the internal material length $c^{1/2}$. Similar results were obtained by Chen et al. (1999) in the theory of phenomenological strain-gradient plasticity (Fleck and Hutchinson, 1997) with the plastic work hardening exponent $n=5$ and by Shi et al. (2000b) in the limit of gradient plasticity for an incompressible material with the plastic work hardening exponent $n=1$. It should be noted however, that there are *several* stress measures involved in gradient elasticity, i.e. the monopolar stresses, the dipolar stresses and the total stresses. Contrary to the classical elasticity, the monopolar stresses (Eqs. (34), (39)) are found to be bounded at the tip of the crack whereas, the dipolar stresses (Eqs. (35), (40)) and the total stresses have a singular character. The important thing to notice, however, is that despite the singular character of these higher order stresses, the $J$-integral is bounded (Georgiadis, 2003; Gourgiotis and Georgiadis, 2009). In particular, it was shown in Gourgiotis and Georgiadis (2009) that when the microstructure of the material is taken into account, the ratio $J/J^{clas.}$, where $J^{clas.}$ is the expression of the $J$-integral in classical elastic fracture mechanics, decreases monotonically with increasing values of $c^{1/2}/a$ ($2a$ being the length of the crack). This finding shows that the gradient theory predicts a *strengthening effect* since a reduction of the crack driving force takes place as the material microstructure becomes more pronounced. An analogous result was found by Shi et al. (2004), where the decrease of the values of $J$ was attributed to 'shielding' dislocations that shield the crack-tip.

(ii) Another important result is that the strain field is *bounded* at the tip of crack. Thus, the necessary condition for uniqueness of the crack problem in form II of Mindlin's gradient elasticity (Grentzelou and Georgiadis, 2005) is fulfilled by the present asymptotic solution. Further, it is noted that the crack faces close more smoothly ($\sim r^{3/2}$) as compared to the classical result. This cusp-like closure has been observed in the experiments by Elssner et al. (1994) and in the results of an analysis through discrete dislocations around a crack tip by Cleveringa et al. (2000).

## 5. Equilibrium considerations for the plane-strain notch problem

In this Section, we proceed to consider the equilibrium of a small circle of radius $r_0$ surrounding the tip of a crack (see Fig. 3), in order to elucidate the role of the concentrated forces $\left(E_r, E_\theta\right)$ defined in (8). The crack faces at $\theta = \pm\pi$ ($\mathbf{n} = \pm\mathbf{e_\theta}$) are traction-free. The distribution of the force tractions $\left(P_r^{(n)}, P_\theta^{(n)}\right)$ and the double-force tractions $\left(R_r^{(n)}, R_\theta^{(n)}\right)$ on the small circle of radius $r_0$ (with $\mathbf{n} = \mathbf{e}_r$) is depicted in Fig. 3. The expressions for tractions are given in Appendix A.

Further, one may observe in Fig. 3 that the outward unit normal $\mathbf{n}$ is discontinuous at the corner points $A(r_0, \pi)$ and $B(r_0, -\pi)$. In particular, as we approach the corner point $A$ ($B$) from the crack face, the outward unit normal is $\mathbf{n} = \mathbf{e}_\theta$ ($\mathbf{n} = -\mathbf{e}_\theta$). As we approach $A$ and $B$ moving along the circle, we have $\mathbf{n} = \mathbf{e}_r$. This discontinuity, according to (8), gives rise to concentrated forces $E_q$ in the corners $A$ and $B$, respectively. As we shall see, the role of these concentrated forces is to *balance* the resultant force and moment of the distributed force and double-force tractions acting on the circle. It should be noted that, in our case, the edge $C$ (defined in Section 2) is a straight line parallel to the $z$-axis and passing through the corner points $A$ ($C_A$) and $B$ ($C_B$), respectively. Therefore, these concentrated forces are line loads (constant upon $z$) acting along the edges $C_A$ and $C_B$.

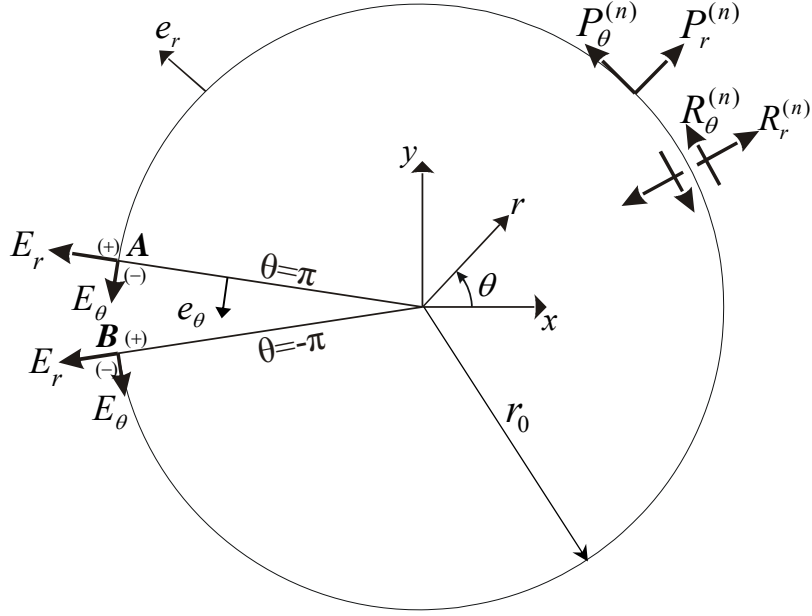

**Fig. 3** Equilibrium of a small circle surrounding the crack tip. The tractions ($P_q^{(n)}$, $R_q^{(n)}$) are distributed along the circle with radius $r_0$.

Moreover, from the geometry of the problem, it is evident that $\mathbf{n}^{(A^+)} = (n_r^{(A^+)}, n_\theta^{(A^+)}) = (1,0)$, $\mathbf{n}^{(A^-)} = (0,1)$, $\mathbf{n}^{(B^+)} = -\mathbf{n}^{(A^-)}$, and $\mathbf{n}^{(B^-)} = \mathbf{n}^{(A^+)}$. Also, the unit tangent vector (in the positive direction), along the edges $C_A$ and $C_B$, is defined as $\mathbf{s} = (0,0,1)$. In view of the above, we may write for the vector $\mathbf{k} = \mathbf{s} \times \mathbf{n}$, the following geometric relations $\mathbf{k}^{(A^+)} = (0,1)$, $\mathbf{k}^{(A^-)} = (-1,0)$, $\mathbf{k}^{(B^+)} = -\mathbf{k}^{(A^-)}$, and $\mathbf{k}^{(B^-)} = \mathbf{k}^{(A^+)}$. It is noted, that the superscript $(+)$ refers to the surface which is to the left of the unit tangent vector $\mathbf{s}$ and $(-)$ to the surface which is to the right (see Fig. 3).

Employing now the definition in (8), the polar components of these concentrated forces at the corner points $A(r_0, \pi)$ and $B(r_0, -\pi)$ read

$$E_r^{(A)} = n_r^{(A^+)} k_\theta^{(A^+)} m_{r\theta r}(r_0, \pi) - n_\theta^{(A^-)} k_r^{(A^-)} m_{\theta rr}(r_0, \pi) = m_{r\theta r}(r_0, \pi) + m_{\theta rr}(r_0, \pi), \qquad (42a)$$

$$E_\theta^{(A)} = n_r^{(A^+)}k_\theta^{(A^+)}m_{r\theta\theta}(r_0,\pi) - n_\theta^{(A^-)}k_r^{(A^-)}m_{\theta r\theta}(r_0,\pi) = m_{r\theta\theta}(r_0,\pi) + m_{\theta r\theta}(r_0,\pi) , \tag{42b}$$

$$E_r^{(B)} = n_\theta^{(B^+)}k_r^{(B^+)}m_{\theta rr}(r_0,-\pi) - n_r^{(B^-)}k_\theta^{(B^-)}m_{r\theta r}(r_0,-\pi) = -m_{\theta rr}(r_0,-\pi) - m_{r\theta r}(r_0,-\pi), \tag{42c}$$

$$E_\theta^{(B)} = n_\theta^{(B^+)}k_r^{(B^+)}m_{\theta r\theta}(r_0,-\pi) - n_r^{(B^-)}k_\theta^{(B^-)}m_{r\theta\theta}(r_0,-\pi) = -m_{\theta r\theta}(r_0,-\pi) - m_{r\theta\theta}(r_0,-\pi) . \tag{42d}$$

In light of the above, the total resultant forces in the horizontal and vertical direction and the total resultant moment with respect to the crack-tip position are written as

$$\Sigma F_x = H - \left[E_r^{(A)} + E_r^{(B)}\right] = r_0 \int_{-\pi}^{\pi} P_r^{(n)} \cos\theta\, d\theta - r_0 \int_{-\pi}^{\pi} P_\theta^{(n)} \sin\theta\, d\theta - \left[E_r^{(A)} + E_r^{(B)}\right]$$

$$= r_0 \int_{-\pi}^{\pi} t_{rr} \cos\theta\, d\theta - r_0 \int_{-\pi}^{\pi} t_{r\theta} \sin\theta\, d\theta - \left[E_r^{(A)} + E_r^{(B)}\right] , \tag{43a}$$

$$\Sigma F_y = V - \left[E_\theta^{(A)} + E_\theta^{(B)}\right] = r_0 \int_{-\pi}^{\pi} P_r^{(n)} \sin\theta\, d\theta + r_0 \int_{-\pi}^{\pi} P_\theta^{(n)} \cos\theta\, d\theta - \left[E_\theta^{(A)} + E_\theta^{(B)}\right]$$

$$= r_0 \int_{-\pi}^{\pi} t_{rr} \sin\theta\, d\theta + r_0 \int_{-\pi}^{\pi} t_{r\theta} \cos\theta\, d\theta - \left[E_\theta^{(A)} + E_\theta^{(B)}\right] , \tag{43b}$$

$$\Sigma M = T + r_0 \left[E_\theta^{(A)} + E_\theta^{(B)}\right] = r_0^2 \int_{-\pi}^{\pi} P_\theta^{(n)}\, d\theta + r_0 \int_{-\pi}^{\pi} R_\theta^{(n)}\, d\theta + r_0 \left[E_\theta^{(A)} + E_\theta^{(B)}\right]$$

$$= r_0^2 \int_{-\pi}^{\pi} t_{r\theta}\, d\theta + r_0 \int_{-\pi}^{\pi} m_{rr\theta}\, d\theta + r_0 \left[E_\theta^{(A)} + E_\theta^{(B)}\right] , \tag{43c}$$

where $(H, V)$ represent the resultant horizontal and vertical forces and $T$ the resultant moment, due to the distributed force and double-force tractions $\left(P_q^{(n)}, R_q^{(n)}\right)$ along the circumference of the circle. Also, $t_{rr}$ and $t_{r\theta}$ are the total stresses that correspond to outward unit normal $\mathbf{n} = \mathbf{e}_r$ (see Eqs. (A8) and (A9) in Appendix A). Further, we note that, in (43c), only the double-force traction $R_\theta^{(n)} \equiv m_{rr\theta}$ contributes in the equilibrium of moments, since $R_r^{(n)} \equiv m_{rrr}$ is a self-equilibrated field (see Fig. 3).

The equilibrium for the mode I (symmetric) case is examined first. It is noted, that since the radius $r_0$ of the circle is very small, only the most singular asymptotic stress fields derived previously will be used to check equilibrium. Furthermore, we can show that $m_{r\theta r}(r_0,\pi) = -m_{r\theta r}(r_0,-\pi),\quad m_{\theta rr}(r_0,\pi) = -m_{\theta rr}(r_0,-\pi),\quad m_{\theta r\theta}(r_0,\pi) = m_{\theta r\theta}(r_0,-\pi) = 0,$ and $m_{r\theta\theta}(r_0,\pi) = m_{r\theta\theta}(r_0,-\pi) = 0$ by invoking the symmetry of the problem and Eqs. (35).

In view of the above and employing Eqs. (42), we finally obtain

$$E_r^{(A)}(r_0,\pi) = m_{r\theta r} + m_{\theta rr} = -\frac{12c\mu r_0^{-1/2}}{(41-32\nu)}[A_1(27-24\nu) + A_2(14-8\nu)] ,$$

$$E_r^{(B)}(r_0,-\pi) = E_r^{(A)}(r_0,\pi) , \qquad E_\theta^{(B)}(r_0,-\pi) = E_\theta^{(A)}(r_0,\pi) = 0 . \tag{44a-c}$$

One may observe, therefore, that in the mode I case only the radial components of the concentrated forces at the corners survive. We also note that these forces are square root singular at the tip of the crack. Further, with the aid of (35) and employing the definitions (A8) and (A9) in Appendix A, we are able to write the total stresses defined on the boundary of the circle $r = r_0$ as

$$t_{rr}(r_0,\theta) = \frac{3c\mu r_0^{-3/2}}{4(41-32\nu)} A_1 \left[(147-32\nu)\cos\frac{3\theta}{2} + (41-32\nu)\cos\frac{\theta}{2}\right]$$

$$+ \frac{3c\mu r_0^{-3/2}}{4(41-32\nu)} A_2 \left[(123-96\nu)\cos\frac{5\theta}{2} - (147-32\nu)\cos\frac{3\theta}{2}\right] , \tag{45}$$

$$t_{r\theta}(r_0,\theta) = \frac{3c\mu r_0^{-3/2}}{4(41-32\nu)} A_1 \left[(43+32\nu)\sin\frac{3\theta}{2} + (451-352\nu)\sin\frac{\theta}{2}\right]$$

$$- \frac{3c\mu r_0^{-3/2}}{4(41-32\nu)} A_2 \left[(123-96\nu)\sin\frac{5\theta}{2} + (43+32\nu)\sin\frac{3\theta}{2}\right] . \tag{46}$$

Moreover, since the total stresses exhibit an $r^{-3/2}$-singularity, we infer that resultant forces $H$ and $V$ (see (43)) exhibit also a square root singularity.

Now, in view of the symmetry of the mode I problem, the vertical resultant force $V$ in (43b) and the resultant moment $T$ in (43c) are identically zero. Thus, we conclude that $\Sigma F_y = 0$ and $\Sigma M = 0$. On the other hand, according to Eqs. (43)-(45), it can readily be shown that the sum of the two concentrated horizontal forces $E_r^{(A)}(r_0,\pi)$ and $E_r^{(B)}(r_0,-\pi)$ balances the resultant horizontal force $H$ due to the distributed force tractions $P_r^{(n)}$ and $P_\theta^{(n)}$ on the circle. Therefore, $\Sigma F_x = 0$ in (43a) and equilibrium prevails in the horizontal direction, as well. It is noted, that as $r_0 \to 0$, the concentrated forces in (44) become infinitely large, resembling a single concentrated force applied at the tip of the crack. However, this force is balanced, as we saw before, by the resultant force $H$, which in this case ($r_0 \to 0$) can also be viewed as a single force acting (in the opposite direction) on the crack tip.

The same observations apply for the *general* notch problem, as well. In particular, in the case of symmetric loading, the equilibrium equations (43b) and (43c) are identically satisfied. Also, it can be shown that the sum of the two concentrated horizontal forces $E_r^{(A)}(r_0, a)$ and $E_r^{(B)}(r_0, -a)$ at the corners, will always be balanced by the resultant horizontal force $H$, due to the distributed force tractions $P_r^{(n)}$ and $P_\theta^{(n)}$ along the circular sector.

In the mode II (anti-symmetric) case, the horizontal resultant force $H$ in (43a) is identically zero. Further, by virtue of Eqs. (40) and (42), the concentrated forces in the corners $A$ and $B$ read

$$E_\theta^{(A)}(r_0, \pi) = m_{r\theta\theta} = \frac{3c\mu r_0^{-1/2}}{2(1-2\nu)}\left[B_1 - B_2 \frac{2(64\nu^2 - 88\nu + 29)}{(37 - 32\nu)}\right],$$

$$E_\theta^{(B)}(r_0, -\pi) = E_\theta^{(A)}(r_0, \pi) , \qquad E_r^{(B)}(r_0, -\pi) = E_r^{(A)}(r_0, \pi) = 0 . \qquad (47\text{a-c})$$

The total stresses for $r = r_0$ become

$$t_{rr}(r_0, \theta) = \frac{3\mu c r_0^{-3/2}}{4(37 - 32\nu)} B_2 \left[\frac{(10 - 28\nu)}{(1 - 2\nu)}\sin\frac{\theta}{2} + 3(37 - 32\nu)\sin\frac{5\theta}{2} - (147 - 32\nu)\sin\frac{3\theta}{2}\right]$$

$$+ \frac{3c\mu r_0^{-3/2}}{2(1 - 2\nu)} B_1 \sin\frac{\theta}{2} , \qquad (48)$$

$$t_{r\theta}(r_0, \theta) = \frac{3\mu c r_0^{-3/2}}{4(37 - 32\nu)} B_2 \left[\frac{(16 - 28\nu)}{(1 - 2\nu)}\cos\frac{\theta}{2} + 3(37 - 32\nu)\cos\frac{5\theta}{2} + (43 + 32\nu)\cos\frac{3\theta}{2}\right]$$

$$- \frac{3c\mu r_0^{-3/2}}{4(1 - 2\nu)} B_1 \cos\frac{\theta}{2} . \qquad (49)$$

In light of the above, it can easily be shown that both the force equilibrium in the vertical sense (42b) and the moment equilibrium (42c) are satisfied.

Finally, we mention that a discussion of these edge forces can be found in Green et al. (1968).

## 6. The limit case of a half space in plane-strain

The limit case of the half-plane ($a = 90^0$) is considered next. The characteristic equation, for both symmetric and antisymmetric loadings, takes the following form

$$(p-1)^4(p-2)^2(1-\cos 2\pi p) = 0 \Rightarrow p = n, \quad n = 0, \pm 1, \pm 2, \dots . \tag{50}$$

According to (50), the first admissible eigenvalue is $p = 2$. Hence, the displacement field for the symmetric case becomes

$$u_r = (C_1 + C_3 \cos 2\theta) r + A_1 r^2 \cos\theta + A_2 r^2 \cos 3\theta , \tag{51a}$$

$$u_\theta = C_3 r \sin 2\theta + A_1 \frac{(\nu - 2)}{\nu} r^2 \sin\theta - A_2 r^2 \left( \frac{2(1-2\nu)}{\nu} \sin\theta + \sin 3\theta \right) . \tag{51b}$$

In light of the above, we conclude that the monopolar stresses exhibit a variation of the type $\sim r$, whereas the dipolar stresses behave as $O(1)$. As for the total stresses, according to Eqs. (B13)-(B14) in Appendix B, the eigenvalue $p = 2$ leads to zero total stresses. To obtain non-zero total stresses, one has to take the next successive eigenvalue, i.e. $p = 3$. This eigenvalue provides a field of total stresses that behaves as $O(1)$ and also gives higher-order terms for the displacements and strains.

**7. The notch problem in anti-plane strain**

We consider a body with a notch (re-entrant corner) occupying a domain in the $(r, \theta)$-plane under conditions of anti-plane strain (see Fig. 4). The following displacement field is then generated

$$u_r = u_\theta = 0, \quad u_z \equiv w \neq 0, \quad w \equiv w(r, \theta) . \tag{52}$$

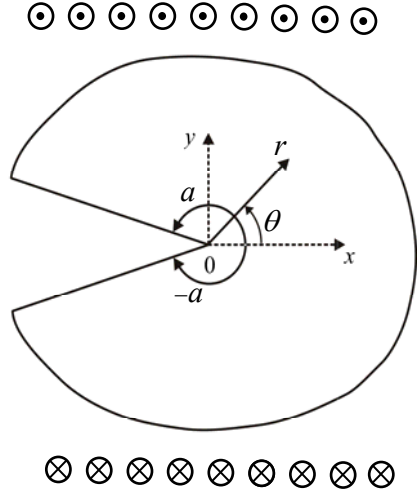

**Fig. 4** The notch problem in anti-plane strain.

The non-vanishing components of the monopolar and dipolar stresses are

$$\tau_{rz} = \mu \frac{\partial w}{\partial r}, \qquad \tau_{\theta z} = \mu \frac{1}{r}\frac{\partial w}{\partial \theta}, \qquad (53a,b)$$

$$m_{rrz} = \mu c \frac{\partial^2 w}{\partial r^2}, \qquad m_{r\theta z} = \mu c \left( \frac{1}{r}\frac{\partial^2 w}{\partial r \partial \theta} - \frac{1}{r^2}\frac{\partial w}{\partial \theta} \right),$$

$$m_{\theta r z} = \mu c \left( \frac{1}{r}\frac{\partial^2 w}{\partial r \partial \theta} - \frac{1}{r^2}\frac{\partial w}{\partial \theta} \right), \qquad m_{\theta\theta z} = \mu c \left( \frac{1}{r}\frac{\partial w}{\partial r} + \frac{1}{r^2}\frac{\partial^2 w}{\partial \theta^2} \right). \qquad (54a\text{-}d)$$

The total stress along the plane $(r, \theta = \text{const.})$ (with $\mathbf{n} = \mathbf{e}_\theta$) is defined as (see Appendix A)

$$P_z^{(n)} \equiv t_{\theta z} = \tau_{\theta z} - \frac{\partial m_{r\theta z}}{\partial r} - \frac{\partial m_{\theta r z}}{\partial r} - \frac{1}{r}\frac{\partial m_{\theta\theta z}}{\partial \theta} - \frac{1}{r} m_{r\theta z} - \frac{1}{r} m_{\theta r z}. \qquad (55)$$

In light of the above, the equation of equilibrium (5), in the case of anti-plane strain, takes the following form

$$c \nabla^4 w - \nabla^2 w = 0. \qquad (56)$$

We focus attention now on the immediate vicinity of the corner point and consider the notch under remotely applied loading. Then, the displacement field takes the following *separated* variable form

$$w(r,\theta) = r^p W(\theta) , \qquad (57)$$

where the exponent $p$ and the angular function $W(\theta)$ are to be determined.

If only the dominant singular terms are retained, the field equation (56) becomes

$$\nabla^4 w = 0 . \qquad (58)$$

The general solution to the biharmonic equation that exhibits an odd (anti-symmetric) behavior in $\theta$, is as follows

$$w = r^p \left[ D_1 \sin p\theta + D_2 \sin(p-2)\theta \right] , \qquad (59)$$

where $D_1$ and $D_2$ are unknown constants.

On the other hand, by retaining only the most singular terms, we write the boundary conditions in terms of displacements under the form

$$t_{\theta z}(r, \pm a) = 0 \Rightarrow 2r^2 \frac{\partial^3 w}{\partial \theta \partial r^2} + 2\frac{\partial w}{\partial \theta} + \frac{\partial^3 w}{\partial \theta^3} - r\frac{\partial^2 w}{\partial r \partial \theta} = 0 , \qquad (60a)$$

$$m_{\theta \theta z}(r, \pm a) = 0 \Rightarrow r\frac{\partial w}{\partial r} + \frac{\partial^2 w}{\partial \theta^2} = 0 . \qquad (60b)$$

Equations (59) and (60) form an *eigenvalue* problem. For a non-trivial solution to exist, the determinant of the coefficients of $(D_1, D_2)$ should vanish and this gives the result

$$(p-1)^2 (p-2)\left[(p-1)\sin 2a + 3\sin 2(p-1)a\right] = 0 . \qquad (61)$$

Further, from the requirement of bounded strain energy in vicinity of the tip of the notch, the exponent $p$ must satisfy the following inequality: $p > 1$. Also, using the same reasoning as in the plane-strain case before, we note that the eigenvalue $p = 1$ should be taken into account since it leads to bounded strain energy. It is noted, however, that this eigenvalue does not contribute to the dipolar stresses (see Eqs. (B20)-(B23) in Appendix B).

In Fig. 5a, the variation of the exponent ($p-1$) of the monopolar stresses in both classical and gradient elasticity (see Eq. (53)) is displayed. We note that, as in the plane-strain case, the monopolar stress is not singular in gradient elasticity. In Fig. 5b, the strength of the singularity ($p-3$) of the total stress $t_{\theta z}$ in (55) is depicted. It is observed that as the angle of the notch $a$ is reduced from $180°$ to $90°$, the strength of the stress singularity falls monotonically from $-1.5$ to $-1$. The strongest singularity corresponds to the case of the crack.

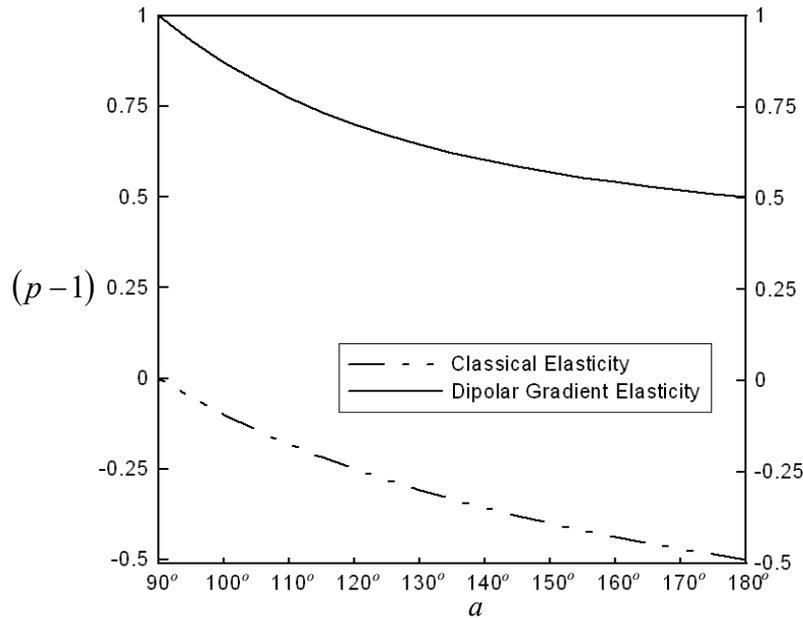

**Fig. 5a** Variation of the monopolar stress with respect to the notch angle $a$ in anti-plane shear.

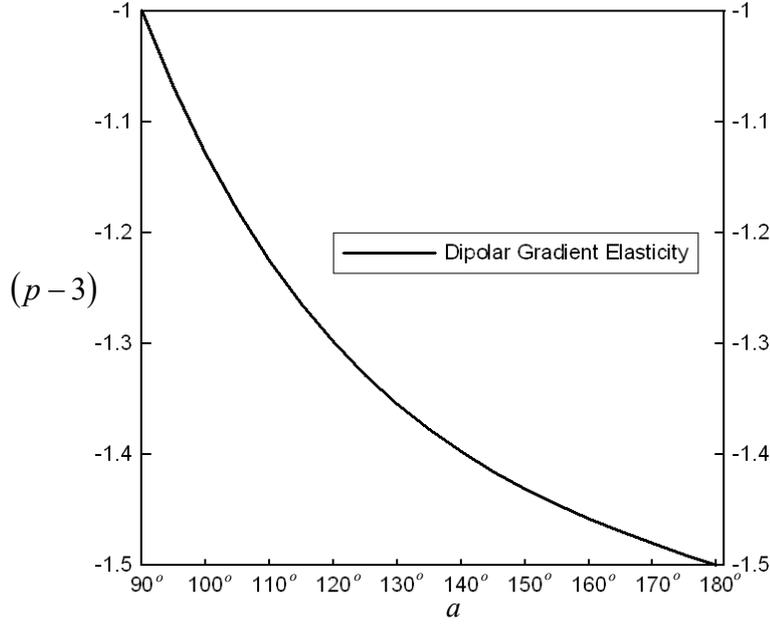

**Fig. 5b** Variation of the exponent of the total stress in gradient elasticity with respect to the notch angle $a$ in anti-plane shear.

Next, the mode III crack is examined as a limit case of a notch with angle $a = 180^0$. It is noted, that this problem was first solved successfully, in the context of dipolar gradient elasticity, by Georgiadis (2003). The characteristic equation (61) now becomes

$$(p-1)^2(p-2)\sin(2p\pi) = 0 \Rightarrow p = n/2 \quad , \quad n = 0, \pm 1, \pm 2,\ldots \quad . \tag{62}$$

The most singular admissible eigenvalue allowed by the energy criterion is $p = 3/2$. The latter eigenvalue gives the most singular solution and, thus, the dominant behavior of the stress field as $r \to 0$. Then, the corresponding displacement field has the form

$$w = Er\sin\theta + (D/3)r^{3/2}\left[5\sin\frac{3\theta}{2} - 3\sin\frac{\theta}{2}\right] \quad , \tag{63}$$

where $E$ is the amplitude factor for the lower-order crack-tip field and $D$ is the amplitude factor for the dominant term of order $3/2$. It is noted, that the linear (in $r$) term does not contribute to the energy release rate as well as to the crack tip opening displacement (Georgiadis, 2003; Radi, 2008). This term only produces a constant strain field at the crack tip (see also Eqs. (B16)-B(17) in Appendix B).

In view of the above, the monopolar and dipolar stresses are written as

$$\tau_{\theta z} = \mu E \cos\theta + \frac{\mu D}{2} r^{1/2}\left[5\cos\frac{3\theta}{2} - \cos\frac{\theta}{2}\right], \tag{64a}$$

$$\tau_{rz} = \mu E \sin\theta + \frac{\mu D}{2} r^{1/2}\left[5\sin\frac{3\theta}{2} - 3\sin\frac{\theta}{2}\right], \tag{64b}$$

$$m_{\theta rz} = \frac{\mu cD}{4} r^{-1/2}\left[5\cos\frac{3\theta}{2} - \cos\frac{\theta}{2}\right], \tag{65a}$$

$$m_{\theta\theta z} = -\frac{5\mu cD}{4} r^{-1/2}\left[\sin\frac{3\theta}{2} + \sin\frac{\theta}{2}\right], \tag{65b}$$

$$m_{rrz} = \frac{\mu cD}{4} r^{-1/2}\left[5\sin\frac{3\theta}{2} - 3\sin\frac{\theta}{2}\right], \tag{65c}$$

$$m_{r\theta z} = m_{\theta rz}, \tag{65d}$$

$$t_{\theta z} = \frac{\mu cD}{8} r^{-3/2}\left[5\cos\frac{3\theta}{2} + 7\cos\frac{\theta}{2}\right]. \tag{66}$$

Finally, we consider the special case of a half space in anti-plane shear ($a = 90^0$). The characteristic equation (61) now becomes

$$(p-1)^2(p-2)\sin(p\pi) = 0 \Rightarrow p = n, \quad n = 0,\pm 1,\pm 2,... \tag{67}$$

The first admissible eigenvalue is $p = 2$. In this case, the displacement field is as follows

$$w = Er\sin\theta + Dr^2\sin(2\theta). \tag{68}$$

The dipolar stresses are given as

$$m_{\theta rz} = 2\mu cD\cos 2\theta, \quad m_{\theta\theta z} = -2\mu cD\sin 2\theta,$$

$$m_{rrz} = 2\mu cD\sin 2\theta, \quad m_{r\theta z} = m_{\theta rz}. \tag{69a-d}$$

It is noted, that according to Eq. (B24) in Appendix B, the eigenvalue $p = 2$ results in zero total stress $t_{\theta z}$. To obtain non-zero total stresses, one has to take the next successive eigenvalue, i.e. $p = 3$. This eigenvalue provides a field of total stresses that behaves as $O(1)$ and also gives higher-order terms for the displacements and strains. The same asymptotic behavior is exhibited by $\tau_{\theta z}$ in classical elasticity in the case of a half space (see Fig. 5a).

## 8. Discussion and concluding remarks

In this paper, the asymptotic displacement, strain and stress fields near the corner of a sharp notch in a body of infinite extent (wedge) are determined by using the theory of dipolar gradient elasticity. Form II of Mindlin's theory was employed to account for effects of microstructure. Plane-strain and anti-plane shear conditions are considered. The notch faces are taken traction-free and the loading is assumed to be remotely applied. The boundary value problem was attacked with the asymptotic Knein-Williams technique. Our analysis leads to an eigenvalue problem, which, along with the restriction of a bounded strain energy, provides the asymptotic fields.

The results for the near-corner fields showed significant departure from the predictions of classical elasticity. In general, the strain field is always *bounded* at the tip of the notch. Also, all asymptotic fields depend not only upon the notch angle $a$, but also upon the Poisson's ratio $\nu$. As for the stresses, there are *several* stress measures in gradient elasticity, i.e. the monopolar stresses, the dipolar stresses and the total stresses. The monopolar stresses are found to be bounded at the tip of the notch, but the other stresses have a singular character.

More specifically, we notice the following points:

(i) According to Eqs. (29) and (30), the displacement field is described in terms of a lower-order, linear (in $r$) term, which produces a constant strain field, and a dominant ($r^p$) term which defines the singular behavior of the dipolar and total stresses. The exponent $p$ varies from $p = 3/2$ (crack case) to $p = 2$ (half-space case). It is noted that, contrary to the classical elasticity case, the linear terms in the displacement field do not vanish for notch angles $90^o < a < 180^o$. This can be justified from the fact that the lower-order linear terms are coupled, through the governing equations (16) and the boundary conditions in (19a,b), with higher-order terms (of $O(r^3)$) in the asymptotic expansion. Recall that, in classical elasticity, these linear terms appear *only* in the symmetric loading case at angles $a = 180^o$ and $a = 90^o$ (Stephen and Wang, 1999).

However, in the mode I and mode II crack problems, these linear terms do not affect the crack opening (Eq. (32b)) or sliding (Eq. (37a)) displacements. Indeed, in these cases, the crack-face displacements exhibit an $r^{3/2}$ variation (cusp-like closure). It is remarked, that this type of closure has also been observed in the experiments by Elssner et al. (1994) and in the results of an analysis through discrete dislocations around a crack tip by Cleveringa et al. (2000). Also, more recently, Xiao and Karihaloo (2006) using the Knein-Williams asymptotic technique, have shown that the crack faces of a pure mode I (frictionless) cohesive crack close in a cusp-like manner.

(ii) The strain field is always bounded at the tip of the notch. In particular, due to the existence of the lower-order linear terms in the displacement field, the strains take a constant (non-zero) value when $r \to 0$ (see also Eqs. (B1)-(B3) in Appendix B). In the case of a crack, they exhibit a variation of the form $\sim \left(\text{const.} + r^{1/2}\right)$. Thus, the necessary condition for uniqueness of the crack problem in form II of Mindlin's gradient elasticity (Grentzelou and Georgiadis, 2005) is fulfilled by the present asymptotic solution.

(iii) The total stresses at the tip of the notch exhibit a stronger singularity than the one predicted by classical elasticity (Fig. 2b). Indeed, in the case of a crack, an aggravation of the stress field, as compared to the respective result of the conventional theory, is observed (this aggravation appears here through the stronger $r^{-3/2}$ singularity). This behavior is in agreement with the analytical results of Shi et al. (2000b), Georgiadis (2003), and Gourgiotis and Georgiadis (2009). Such a strong singularity was also suggested by the experimental evidence of Prakash et al. (1992) in extremely brittle fracture. Notice, however, that despite the singular character of these higher order stresses, the $J$-integral remains bounded (Georgiadis, 2003; Gourgiotis and Georgiadis, 2009). In particular, it was shown in Gourgiotis and Georgiadis (2009) that when the microstructure of the material is taken into account, the ratio $J/J^{clas.}$, where $J^{clas.}$ is the expression of the $J$-integral in classical elastic fracture mechanics, decreases monotonically with increasing values of $c^{1/2}/a$ ($2a$ being the length of the crack). This finding shows that the gradient theory predicts a *strengthening effect* since a reduction of the crack driving force takes place as the material microstructure becomes more pronounced.

**Appendix A: Boundary conditions and equilibrium equations in polar coordinates**

In this Appendix, we derive the total stresses and the equilibrium equations in polar coordinates, in the case of plane and antiplane strain.

The boundary condition (6) can be written in *direct* form as

$$\mathbf{P}^{(n)} = \mathbf{n} \cdot (\boldsymbol{\tau} - \nabla \cdot \mathbf{m}) - \overset{s}{\nabla} \cdot (\mathbf{n} \cdot \mathbf{m}) + (\overset{s}{\nabla} \cdot \mathbf{n})\mathbf{n} \cdot (\mathbf{n} \cdot \mathbf{m}) \,, \tag{A1}$$

where $\overset{s}{\nabla} = (\mathbf{I} - \mathbf{nn}) \cdot \nabla$ is the surface gradient operator, $\mathbf{I}$ is the unit dyadic and $\nabla$ is the usual gradient operator given as $\nabla() = \mathbf{e}_r \partial_r() + \mathbf{e}_\theta r^{-1} \partial_\theta()$ in polar coordinates. Also, we note that the base vectors are related through the following differential relations $\partial_\theta \mathbf{e}_r = \mathbf{e}_\theta$, $\partial_\theta \mathbf{e}_\theta = -\mathbf{e}_r$, $\partial_r \mathbf{e}_r = 0$, and $\partial_r \mathbf{e}_\theta = 0$.

The cases of plane and anti-plane strain are examined now separately.

*Plane strain*

The monopolar and dipolar stress tensors, in the case of plane strain, are written as

$$\boldsymbol{\tau} = \tau_{rr}\, \mathbf{e}_r \otimes \mathbf{e}_r + \tau_{\theta r}\, \mathbf{e}_\theta \otimes \mathbf{e}_r + \tau_{r\theta}\, \mathbf{e}_r \otimes \mathbf{e}_\theta + \tau_{\theta\theta}\, \mathbf{e}_\theta \otimes \mathbf{e}_\theta + \tau_{zz}\, \mathbf{e}_z \otimes \mathbf{e}_z \,, \tag{A2}$$

$$\begin{aligned}\mathbf{m} =\ & m_{rrr}\, \mathbf{e}_r \otimes \mathbf{e}_r \otimes \mathbf{e}_r + m_{r\theta r}\, \mathbf{e}_r \otimes \mathbf{e}_\theta \otimes \mathbf{e}_r + m_{rr\theta}\, \mathbf{e}_r \otimes \mathbf{e}_r \otimes \mathbf{e}_\theta + m_{r\theta\theta}\, \mathbf{e}_r \otimes \mathbf{e}_\theta \otimes \mathbf{e}_\theta \\ & + m_{\theta\theta\theta}\, \mathbf{e}_\theta \otimes \mathbf{e}_\theta \otimes \mathbf{e}_\theta + m_{\theta rr}\, \mathbf{e}_\theta \otimes \mathbf{e}_r \otimes \mathbf{e}_r + m_{\theta r\theta}\, \mathbf{e}_\theta \otimes \mathbf{e}_r \otimes \mathbf{e}_\theta + m_{\theta\theta r}\, \mathbf{e}_\theta \otimes \mathbf{e}_\theta \otimes \mathbf{e}_r \\ & + m_{\theta zz}\, \mathbf{e}_\theta \otimes \mathbf{e}_z \otimes \mathbf{e}_z + m_{zrz}\, \mathbf{e}_z \otimes \mathbf{e}_r \otimes \mathbf{e}_z + m_{rzz}\, \mathbf{e}_r \otimes \mathbf{e}_z \otimes \mathbf{e}_z + m_{zzr}\, \mathbf{e}_z \otimes \mathbf{e}_z \otimes \mathbf{e}_r \\ & + m_{z\theta z}\, \mathbf{e}_z \otimes \mathbf{e}_\theta \otimes \mathbf{e}_z + m_{zz\theta}\, \mathbf{e}_z \otimes \mathbf{e}_z \otimes \mathbf{e}_\theta \,. \end{aligned} \tag{A3}$$

Further, when $\mathbf{n} = \mathbf{e}_\theta$, the surface gradient operator takes the form $\overset{s}{\nabla}() = \mathbf{e}_r \partial_r()$. In this case we obtain

$$\mathbf{n} \cdot \mathbf{m} = m_{\theta rr}\, \mathbf{e}_r \otimes \mathbf{e}_r + m_{\theta r\theta}\, \mathbf{e}_r \otimes \mathbf{e}_\theta + m_{\theta\theta r}\, \mathbf{e}_\theta \otimes \mathbf{e}_r + m_{\theta\theta\theta}\, \mathbf{e}_\theta \otimes \mathbf{e}_\theta + m_{\theta zz}\, \mathbf{e}_z \otimes \mathbf{e}_z \,, \tag{A4}$$

$$\mathbf{n} \cdot (\nabla \cdot \mathbf{m}) = \left[ \partial_r m_{r\theta r} + \frac{1}{r} \partial_\theta m_{\theta\theta r} + \frac{1}{r} m_{\theta r r} - \frac{1}{r} m_{\theta\theta\theta} + \frac{1}{r} m_{r\theta r} \right] \mathbf{e}_r$$

$$+ \left[ \partial_r m_{r\theta\theta} + \frac{1}{r} \partial_\theta m_{\theta\theta\theta} + \frac{1}{r} m_{r\theta\theta} + \frac{1}{r} m_{\theta r\theta} + \frac{1}{r} m_{\theta\theta r} \right] \mathbf{e}_\theta , \qquad (A5)$$

$$\overset{s}{\nabla} \cdot (\mathbf{n} \cdot \mathbf{m}) = \partial_r m_{\theta r r} \mathbf{e}_r + \partial_r m_{\theta r\theta} \mathbf{e}_\theta , \qquad (A6)$$

$$\overset{s}{\nabla} \cdot \mathbf{n} = 0 . \qquad (A7)$$

In view of the above, we are able to write for the total stresses Eqs. (14) and (15) of the main text.

On the other hand, when the boundary is defined by $\mathbf{n} = \mathbf{e}_r$, the pertinent total and dipolar stresses along the plane $(r = \text{const.}, \theta)$ become (see also Eshel and Rosenfeld, 1970; 1975)

$$P_r^{(n)} \equiv t_{rr} = \tau_{rr} - \frac{\partial m_{rrr}}{\partial r} - \frac{1}{r} \frac{\partial m_{\theta rr}}{\partial \theta} - \frac{1}{r} \frac{\partial m_{r\theta r}}{\partial \theta} - \frac{1}{r} m_{rrr} + \frac{2}{r} m_{\theta\theta r} + \frac{1}{r} m_{r\theta\theta} , \qquad (A8)$$

$$P_\theta^{(n)} \equiv t_{r\theta} = \tau_{r\theta} - \frac{\partial m_{rr\theta}}{\partial r} - \frac{1}{r} \frac{\partial m_{r\theta\theta}}{\partial \theta} - \frac{1}{r} \frac{\partial m_{\theta r\theta}}{\partial \theta} - \frac{1}{r} m_{\theta rr} - \frac{2}{r} m_{rr\theta} + \frac{1}{r} m_{\theta\theta\theta} . \qquad (A9)$$

$$R_r^{(n)} \equiv m_{rrr} , \quad R_\theta^{(n)} \equiv m_{rr\theta} . \qquad (A10a,b)$$

It should be noted that the total stresses $(t_{rr}, t_{r\theta})$ and $(t_{\theta r}, t_{\theta\theta})$ which correspond to $\mathbf{n} = \mathbf{e}_r$ and $\mathbf{n} = \mathbf{e}_\theta$ respectively, do *not* constitute components of a tensor.

As for the equations of equilibrium in terms of displacements, these are written in direct form as

$$(1 - c\nabla^2)\left[(1 - 2\nu)\nabla \cdot (\nabla \mathbf{u}) + \nabla(\nabla \cdot \mathbf{u})\right] = 0 . \qquad (A11)$$

It is noted that the equations in brackets are the Navier-Cauchy equations of classical elasticity. Now, (A11) can be written in a more convenient form as

$$(1 - c\nabla^2)[s_r \mathbf{e}_r + s_\theta \mathbf{e}_\theta] = 0 \Rightarrow$$

$$\Rightarrow \left[s_r - c[\nabla^2 s_r - r^{-2} s_r - 2r^{-2} \partial_\theta s_\theta]\right] \mathbf{e}_r + \left[s_\theta - c[\nabla^2 s_\theta - r^{-2} s_\theta + 2r^{-2} \partial_\theta s_r]\right] \mathbf{e}_\theta = 0 . \qquad (A12)$$

*Antiplane strain*

The monopolar and dipolar stress tensors in this case are defined as

$$\boldsymbol{\tau} = \tau_{rz}\mathbf{e}_r \otimes \mathbf{e}_z + \tau_{\theta z}\mathbf{e}_\theta \otimes \mathbf{e}_z + \tau_{zr}\mathbf{e}_z \otimes \mathbf{e}_r + \tau_{z\theta}\mathbf{e}_z \otimes \mathbf{e}_\theta ,\qquad(A13)$$

$$\mathbf{m} = m_{rrz}\mathbf{e}_r \otimes \mathbf{e}_r \otimes \mathbf{e}_z + m_{r\theta z}\mathbf{e}_r \otimes \mathbf{e}_\theta \otimes \mathbf{e}_z + m_{\theta rz}\mathbf{e}_\theta \otimes \mathbf{e}_r \otimes \mathbf{e}_z + m_{\theta\theta z}\mathbf{e}_\theta \otimes \mathbf{e}_\theta \otimes \mathbf{e}_z$$
$$+ m_{rzr}\mathbf{e}_r \otimes \mathbf{e}_z \otimes \mathbf{e}_r + m_{rz\theta}\mathbf{e}_r \otimes \mathbf{e}_z \otimes \mathbf{e}_\theta + m_{\theta zr}\mathbf{e}_\theta \otimes \mathbf{e}_z \otimes \mathbf{e}_r + m_{\theta z\theta}\mathbf{e}_\theta \otimes \mathbf{e}_z \otimes \mathbf{e}_\theta .\qquad(A14)$$

Further, when $\mathbf{n} = \mathbf{e}_\theta$, the following relations hold

$$\mathbf{n}\cdot\mathbf{m} = m_{\theta r}\mathbf{e}_r \otimes \mathbf{e}_z + m_{\theta\theta z}\mathbf{e}_\theta \otimes \mathbf{e}_z + m_{\theta zr}\mathbf{e}_z \otimes \mathbf{e}_r + m_{\theta z\theta}\mathbf{e}_z \otimes \mathbf{e}_\theta ,\qquad(A15)$$

$$\mathbf{n}\cdot(\nabla\cdot\mathbf{m}) = [\partial_r m_{rz\theta} + \frac{1}{r}m_{r\theta z} + \frac{1}{r}m_{\theta zr} + \frac{1}{r}\partial_\theta m_{\theta\theta z}]\mathbf{e}_z ,\qquad(A16)$$

$$\overset{s}{\nabla}\cdot(\mathbf{n}\cdot\mathbf{m}) = \partial_r m_{\theta zr}\mathbf{e}_z ,\qquad(A17)$$

$$\overset{s}{\nabla}\cdot\mathbf{n} = 0 .\qquad(A18)$$

In view of the above, we are able to write for the total stress Eq. (52) of the main text.

**Appendix B: General expressions for the strain and stress fields**

*Plane strain*

By virtue of Eqs. (29), (30) and appropriate definitions in the previous analysis, the strains and the monopolar stresses are written as

$$\varepsilon_{rr} = \begin{Bmatrix} C_1 + C_3\cos 2\theta \\ C_2 \sin 2\theta \end{Bmatrix} + pr^{p-1}\left[\begin{Bmatrix} A_1\cos(p-1)\theta \\ B_1\sin(p-1)\theta \end{Bmatrix} + \begin{Bmatrix} A_2\cos(p+1)\theta \\ B_2\sin(p+1)\theta \end{Bmatrix} + \begin{Bmatrix} A_3\cos(p-3)\theta \\ B_3\sin(p-3)\theta \end{Bmatrix}\right],$$
(B1)

$$\varepsilon_{\theta\theta} = \begin{Bmatrix} C_1 - C_3 \cos 2\theta \\ -C_2 \sin 2\theta \end{Bmatrix} + r^{p-1} \left[ \begin{Bmatrix} A_1 \cos(p-1)\theta \\ B_1 \sin(p-1)\theta \end{Bmatrix} - p \begin{Bmatrix} A_2 \cos(p+1)\theta \\ B_2 \sin(p+1)\theta \end{Bmatrix} \right.$$

$$\left. - \frac{p^2 + p - 8p\nu + 16\nu - 8}{p + 8\nu - 7} \begin{Bmatrix} A_3 \cos(p-3)\theta \\ B_3 \sin(p-3)\theta \end{Bmatrix} + (p-1) \begin{Bmatrix} A_4 \cos(p-1)\theta \\ -B_4 \sin(p-1)\theta \end{Bmatrix} \right], \quad \text{(B2)}$$

$$\varepsilon_{\theta r} = \begin{Bmatrix} -C_3 \sin 2\theta \\ C_2 \cos 2\theta \end{Bmatrix} + r^{p-1} \left\{ \frac{(p-1)}{2} \begin{Bmatrix} -A_1 \sin(p-1)\theta \\ B_1 \cos(p-1)\theta \end{Bmatrix} + p \begin{Bmatrix} -A_2 \sin(p+1)\theta \\ B_2 \cos(p+1)\theta \end{Bmatrix} \right.$$

$$\left. + \frac{p^2 - 3p - 8\nu + 8}{p + 8\nu - 7} \begin{Bmatrix} -A_3 \sin(p-3)\theta \\ B_3 \cos(p-3)\theta \end{Bmatrix} + \frac{(p-1)}{2} \begin{Bmatrix} A_4 \sin(p-1)\theta \\ B_4 \cos(p-1)\theta \end{Bmatrix} \right\}. \quad \text{(B3)}$$

$$\tau_{rr} = 2\mu \begin{Bmatrix} C_1/(1-2\nu) + C_3 \cos 2\theta \\ C_2 \sin 2\theta \end{Bmatrix} + 2\mu r^{p-1} \left[ \frac{p + \nu - p\nu}{1 - 2\nu} \begin{Bmatrix} A_1 \cos(p-1)\theta \\ B_1 \sin(p-1)\theta \end{Bmatrix} + p \begin{Bmatrix} A_2 \cos(p+1)\theta \\ B_2 \sin(p+1)\theta \end{Bmatrix} \right.$$

$$\left. + \frac{p^2 - 7p + 8\nu}{p + 8\nu - 7} \begin{Bmatrix} A_3 \cos(p-3)\theta \\ B_3 \sin(p-3)\theta \end{Bmatrix} + \frac{\nu(p-1)}{1 - 2\nu} \begin{Bmatrix} A_4 \cos(p-1)\theta \\ -B_4 \sin(p-1)\theta \end{Bmatrix} \right], \quad \text{(B4)}$$

$$\tau_{\theta\theta} = 2\mu \begin{Bmatrix} C_1/(1-2\nu) - C_3 \cos 2\theta \\ -C_2 \sin 2\theta \end{Bmatrix} + 2\mu r^{p-1} \left[ \frac{1 - \nu + p\nu}{1 - 2\nu} \begin{Bmatrix} A_1 \cos(p-1)\theta \\ B_1 \sin(p-1)\theta \end{Bmatrix} - p \begin{Bmatrix} A_2 \cos(p+1)\theta \\ B_2 \sin(p+1)\theta \end{Bmatrix} \right.$$

$$\left. - \frac{p^2 + p + 8\nu - 8}{p + 8\nu - 7} \begin{Bmatrix} A_3 \cos(p-3)\theta \\ B_3 \sin(p-3)\theta \end{Bmatrix} + \frac{1 - \nu}{1 - 2\nu}(p-1) \begin{Bmatrix} A_4 \cos(p-1)\theta \\ -B_4 \sin(p-1)\theta \end{Bmatrix} \right], \quad \text{(B5)}$$

$$\tau_{\theta r} = \tau_{r\theta} = 2\mu \begin{Bmatrix} -C_3 \sin 2\theta \\ C_2 \cos 2\theta \end{Bmatrix} + \mu r^{p-1} \left\{ (p-1) \begin{Bmatrix} -A_1 \sin(p-1)\theta \\ B_1 \cos(p-1)\theta \end{Bmatrix} + 2p \begin{Bmatrix} -A_2 \sin(p+1)\theta \\ B_2 \cos(p+1)\theta \end{Bmatrix} \right.$$

$$\left. + 2\frac{(p^2 - 3p - 8\nu + 8)}{p + 8\nu - 7} \begin{Bmatrix} -A_3 \sin(p-3)\theta \\ B_3 \cos(p-3)\theta \end{Bmatrix} + (p-1) \begin{Bmatrix} A_4 \sin(p-1)\theta \\ B_4 \cos(p-1)\theta \end{Bmatrix} \right\}. \quad \text{(B6)}$$

Moreover, the dipolar stresses become

$$m_{rrr} = 2c\mu(p-1)r^{p-2} \left[ \frac{p + \nu - p\nu}{1 - 2\nu} \begin{Bmatrix} A_1 \cos(p-1)\theta \\ B_1 \sin(p-1)\theta \end{Bmatrix} + p \begin{Bmatrix} A_2 \cos(p+1)\theta \\ B_2 \sin(p+1)\theta \end{Bmatrix} \right.$$

$$+\frac{p^2-7p+8\nu}{p+8\nu-7}\begin{Bmatrix}A_3\cos(p-3)\theta\\B_3\sin(p-3)\theta\end{Bmatrix}+\frac{\nu(p-1)}{1-2\nu}\begin{Bmatrix}A_4\cos(p-1)\theta\\-B_4\sin(p-1)\theta\end{Bmatrix}\Bigg], \quad \text{(B7)}$$

$$m_{\theta\theta\theta}=2c\mu(p-1)r^{p-2}\Bigg[\frac{3\nu-2-p\nu}{1-2\nu}\begin{Bmatrix}A_1\sin(p-1)\theta\\-B_1\cos(p-1)\theta\end{Bmatrix}+p\begin{Bmatrix}A_2\sin(p+1)\theta\\-B_2\cos(p+1)\theta\end{Bmatrix}$$

$$+\frac{p^2-3p+8\nu-8}{p+8\nu-7}\begin{Bmatrix}A_3\sin(p-3)\theta\\-B_3\cos(p-3)\theta\end{Bmatrix}+\frac{2-p-3\nu+p\nu}{1-2\nu}\begin{Bmatrix}A_4\sin(p-1)\theta\\B_4\cos(p-1)\theta\end{Bmatrix}\Bigg],$$

(B8)

$$m_{\theta r\theta}=-c\mu(p-1)r^{p-2}\Bigg[(p-3)\begin{Bmatrix}A_1\cos(p-1)\theta\\B_1\sin(p-1)\theta\end{Bmatrix}+2p\begin{Bmatrix}A_2\cos(p+1)\theta\\B_2\sin(p+1)\theta\end{Bmatrix}$$

$$+2\frac{(p^2-7p-8\nu+16)}{p+8\nu-7}\begin{Bmatrix}A_3\cos(p-3)\theta\\B_3\sin(p-3)\theta\end{Bmatrix}+(p-3)\begin{Bmatrix}-A_4\cos(p-1)\theta\\B_4\sin(p-1)\theta\end{Bmatrix}\Bigg], \quad \text{(B9)}$$

$$m_{r\theta r}=c\mu(p-1)r^{p-2}\Bigg[(p-1)\begin{Bmatrix}-A_1\sin(p-1)\theta\\B_1\cos(p-1)\theta\end{Bmatrix}+2p\begin{Bmatrix}-A_2\sin(p+1)\theta\\B_2\cos(p+1)\theta\end{Bmatrix}$$

$$+2\frac{(p^2-3p-8\nu+8)}{p+8\nu-7}\begin{Bmatrix}-A_3\sin(p-3)\theta\\B_3\cos(p-3)\theta\end{Bmatrix}+(p-1)\begin{Bmatrix}A_4\sin(p-1)\theta\\B_4\cos(p-1)\theta\end{Bmatrix}\Bigg], \quad \text{(B10)}$$

$$m_{\theta rr}=2c\mu(p-1)r^{p-2}\Bigg[\frac{1-3\nu-p+p\nu}{1-2\nu}\begin{Bmatrix}A_1\sin(p-1)\theta\\-B_1\cos(p-1)\theta\end{Bmatrix}+p\begin{Bmatrix}-A_2\sin(p+1)\theta\\B_2\cos(p+1)\theta\end{Bmatrix}$$

$$+\frac{p^2-11p+8\nu+16}{p+8\nu-7}\begin{Bmatrix}-A_3\sin(p-3)\theta\\B_3\cos(p-3)\theta\end{Bmatrix}-\frac{1-3\nu+p\nu}{1-2\nu}\begin{Bmatrix}A_4\sin(p-1)\theta\\B_4\cos(p-1)\theta\end{Bmatrix}\Bigg],$$

(B11)

$$m_{r\theta\theta}=2c\mu(p-1)r^{p-2}\Bigg[\frac{1-\nu+p\nu}{1-2\nu}\begin{Bmatrix}A_1\cos(p-1)\theta\\B_1\sin(p-1)\theta\end{Bmatrix}-p\begin{Bmatrix}A_2\cos(p+1)\theta\\B_2\sin(p+1)\theta\end{Bmatrix}$$

$$-\frac{p^2+p+8\nu-8}{p+8\nu-7}\begin{Bmatrix}A_3\cos(p-3)\theta\\B_3\sin(p-3)\theta\end{Bmatrix}+\frac{(1-\nu)(p-1)}{1-2\nu}\begin{Bmatrix}A_4\cos(p-1)\theta\\-B_4\sin(p-1)\theta\end{Bmatrix}\Bigg]. \quad \text{(B12)}$$

Accordingly, the total stresses can be written as

$$t_{\theta r} = 2c\mu(p-1)(p-2)r^{p-3}\left[\frac{(1+p)(1-v)}{1-2v}\begin{Bmatrix} A_1\sin(p-1)\theta \\ -B_1\cos(p-1)\theta \end{Bmatrix} + p\begin{Bmatrix} A_2\sin(p+1)\theta \\ -B_2\cos(p+1)\theta \end{Bmatrix}\right.$$

$$\left.+\frac{p^2-3p+8v-8}{p+8v-7}\begin{Bmatrix} A_3\sin(p-3)\theta \\ -B_3\cos(p-3)\theta \end{Bmatrix} + \frac{v-1+pv}{1-2v}\begin{Bmatrix} A_4\sin(p-1)\theta \\ B_4\cos(p-1)\theta \end{Bmatrix}\right], \quad \text{(B13)}$$

$$t_{\theta\theta} = c\mu(p-1)(p-2)r^{p-3}\left[(p+1)\begin{Bmatrix} A_1\cos(p-1)\theta \\ B_1\sin(p-1)\theta \end{Bmatrix} + 2p\begin{Bmatrix} A_2\cos(p+1)\theta \\ B_2\sin(p+1)\theta \end{Bmatrix}\right.$$

$$\left.+2\frac{(p^2+p-8v+8)}{p+8v-7}\begin{Bmatrix} A_3\cos(p-3)\theta \\ B_3\sin(p-3)\theta \end{Bmatrix} + (p+1)\begin{Bmatrix} -A_4\cos(p-1)\theta \\ B_4\sin(p-1)\theta \end{Bmatrix}\right]. \quad \text{(B14)}$$

*Antiplane strain*

In this case, the displacement and the strains take the following form

$$w = Er\sin\theta + r^p(D_1\sin p\theta + D_2\sin(p-2)\theta), \tag{B15}$$

$$\varepsilon_{rz} = \frac{E}{2}\sin\theta + r^{p-1}\frac{p}{2}[D_1\sin p\theta + D_2\sin(p-2)\theta], \tag{B16}$$

$$\varepsilon_{\theta z} = \frac{E}{2}\cos\theta + r^{p-1}\frac{1}{2}[D_1 p\cos p\theta + D_2(p-2)\cos(p-2)\theta]. \tag{B17}$$

Further, by virtue of Eqs. (53)-(55), the general forms of the monopolar, dipolar and total stresses are as follows

$$\tau_{\theta z} = \mu E\cos\theta + \mu r^{p-1}[D_1 p\cos p\theta + D_2(p-2)\cos(p-2)\theta], \tag{B18}$$

$$\tau_{rz} = \mu E\sin\theta + \mu r^{p-1}[D_1 p\sin p\theta + D_2 p\sin(p-2)\theta], \tag{B19}$$

$$m_{\theta rz} = \mu c r^{p-2}(p-1)[D_1 p\cos p\theta + D_2(p-2)\cos(p-2)\theta], \tag{B20}$$

$$m_{r\theta z} = \mu c r^{p-2}(p-1)[D_1 p\cos p\theta + D_2(p-2)\cos(p-2)\theta], \tag{B21}$$

$$m_{\theta\theta z} = -\mu c r^{p-2}(p-1)[D_1 p\sin p\theta + D_2(p-4)\sin(p-2)\theta], \tag{B22}$$

$$m_{rrz} = \mu c r^{p-2} p(p-1)\left[D_1 \sin p\theta + D_2 \sin(p-2)\theta\right] . \tag{B23}$$

$$t_{\theta z} = -\mu c r^{p-3}(p-1)(p-2)\left[D_1 p \cos(p\theta) + D_2(p+2)\cos((p-2)\theta)\right] , \tag{B24}$$

**Acknowledgement:** The authors acknowledge with thanks support from the 'ΠΕΒΕ 2008' programme of NTU Athens (# 65/1695, title of the individual project: 'Fracture, contact and wave propagation problems in dipolar gradient elasticity').